\documentclass[a4paper,twocolumn,11pt,accepted=2024-12-12]{quantumarticle}
\pdfoutput=1
\usepackage[utf8]{inputenc}
\usepackage{graphicx}
\usepackage{float}
\usepackage{amssymb}
\usepackage{amsmath}  
\usepackage{adjustbox}
\usepackage{mathtools}
\usepackage{dsfont}
\usepackage{array}
\usepackage{bm,fixmath}
\usepackage{mathrsfs}
\usepackage{pifont}
\usepackage{multirow}
\usepackage{upgreek}
\usepackage{xcolor}
\usepackage{bm}
\usepackage{bbm}
\usepackage{physics}
\usepackage{comment}
\usepackage{tikz}
\usetikzlibrary{quantikz}
\usepackage{color, soul}
\usepackage{placeins}

\usepackage[pdftex,
            pdftitle={Scattering Free Fermions},
            pdfauthor={Yahui Chai, Arianna Crippa, Karl Jansen, Stefan Kühn, Vincent R. Pascuzzi,Francesco Tacchino, Ivano Tavernelli},
            bookmarks,
            colorlinks,
            linkcolor=myblue,
            citecolor=mymagenta,
            menucolor=black,
            urlcolor=myblue,
            plainpages=false,
            pdfpagelabels,
            hypertexnames=false]{hyperref}

\graphicspath{{figures/}}

\definecolor{mymagenta}{RGB}{200, 0, 100}
\definecolor{myblue}{RGB}{45, 48, 146}
\definecolor{mypurple}{RGB}{200, 112, 255}

\begin{document}

\title{Fermionic wave packet scattering: a quantum computing approach}

\author{Yahui Chai}
\email{yahui.chai@desy.de}
\affiliation{Deutsches Elektronen-Synchrotron DESY, Platanenallee 6, 15738 Zeuthen, Germany}

\author{Arianna Crippa}
\affiliation{Deutsches Elektronen-Synchrotron DESY, Platanenallee 6, 15738 Zeuthen, Germany}
\affiliation{Institut für Physik, Humboldt-Universität zu Berlin, Newtonstr. 15, 12489 Berlin, Germany}

\author{Karl Jansen}
\affiliation{Deutsches Elektronen-Synchrotron DESY, Platanenallee 6, 15738 Zeuthen, Germany}
\affiliation{Computation-Based Science and Technology Research Center, The Cyprus Institute, 20 Kavafi Street,
2121 Nicosia, Cyprus}

\author{Stefan K\"uhn}
\affiliation{Deutsches Elektronen-Synchrotron DESY, Platanenallee 6, 15738 Zeuthen, Germany}

\author{Vincent R.\ Pascuzzi}
\email{vrpascuzzi@ibm.com}
\affiliation{IBM Quantum, IBM Thomas J.\ Watson Research Center, Yorktown Heights, NY 10598, USA}

\author{Francesco Tacchino}
\affiliation{IBM Quantum, IBM Research Europe – Zurich, Rueschlikon 8803, Switzerland}

\author{Ivano Tavernelli}
\affiliation{IBM Quantum, IBM Research Europe – Zurich, Rueschlikon 8803, Switzerland}

\date{(Dated: December 12, 2024)}

\begin{abstract}
    Quantum computing provides a novel avenue towards simulating dynamical phenomena, and, in particular, scattering processes relevant for exploring the structure of matter. However, preparing and evolving particle wave packets on a quantum device is a nontrivial task. In this work, we propose a method to prepare Gaussian wave packets with momentum on top of the interacting ground state of a fermionic Hamiltonian. Using Givens rotation, we show how to efficiently obtain expectation values of observables throughout the evolution of the wave packets on digital quantum computers. We demonstrate our technique by applying it to the staggered lattice formulation of the Thirring model and studying the scattering of two wave packets. Monitoring the particle density and the entropy produced during the scattering process, we characterize the phenomenon and provide a first step towards studying more complicated collision processes on digital quantum computers. In addition, we perform a small-scale demonstration on IBM's quantum hardware, showing that our method is suitable for current and near-term quantum devices.
\end{abstract}

\maketitle

\section{Introduction\label{sec:intro}}
Collider experiments play a central role for understanding the subatomic structure of matter, and for developing and verifying the theoretical description of the elementary particles and their interactions.
Arguably, the most prominent example is the LHC at CERN, which enabled a major experimental breakthrough with the discovery of the Higgs boson~\cite{ATLAS2012a,CMS2012a}.
While the Standard Model provides a theoretical description for all known elementary particles and the fundamental forces among them (except for gravity) in terms of gauge theories, the understanding of scattering processes and more general out-of-equilibrium dynamics at a fundamental level remains challenging.
In particular, these phenomena are nonperturbative, and a standard tool for exploring this regime is lattice field theory.
Discretizing a theory on a lattice allows for sophisticated Markov chain Monte Carlo simulations that have been highly successful for exploring static properties of gauge theories~\cite{Durr2008, Alexandrou_2014}.
However, the sign problem, among others, prevents an extension of this approach to dynamical problems, leaving open many relevant questions regarding the real-time evolution of scattering processes. Hence, there is an ongoing effort to find alternative approaches allowing for overcoming this limitation~\cite{Hebenstreit2013a, Hebenstreit2013, Banuls2019, Banuls2019a, Bauer_2023, beck2023quantuminformationsciencetechnology, dimeglio2023quantumcomputinghighenergyphysics}.

Methods originating from quantum information theory provide a promising avenue towards studying real-time dynamics of lattice field theories.
For instance, classical computational methods based on Tensor Networks have been demonstrated to allow for reliable simulations in regimes where conventional Monte Carlo methods suffer from the sign problem~\cite{Banuls2018a, Banuls2019}.
Tensor Network states are a family of entanglement-based ansätze for the wave function of a strongly-correlated quantum many-body system, which allow for an efficient classical description of the system in scenarios where the entanglement in the system is not too large~\cite{Orus2014a,Bridgeman2017,Banuls2022}.
In particular, Tensor Network states were successfully used to study real-time dynamics of various (1+1)-dimensional Abelian and non-Abelian lattice gauge theories~\cite{Buyens2016b, Kuehn2015}, including meson scattering in the Schwinger model~\cite{Simone_prd}.
So far, Tensor Networks have been most successful for (1+1)-dimensional theories.
Generalizing this success to higher dimensions is not an immediate task, although there already exist calculations for gauge models in higher dimensions~\cite{Felser2019, Magnifico2021}.
Moreover, the Tensor Network approach becomes inefficient in situations where the system becomes highly entangled.
A prominent example are out-of-equilibrium dynamics following a global quench, during which time the entanglement in the system can grow linearly in time~\cite{Calabrese2005, Schuch2008, Schachenmayer2013}.
Hence, for these problems only short to intermediate time-scales are accessible with Tensor Networks~\cite{Banuls2009, Trotzky2012}.

An emerging technology that aims to address shortcomings of classical information processing is quantum computing, which in principle can efficiently simulate real-time dynamics of lattice field theories, even in regimes that are challenging for Tensor Networks.
Indeed, there have already been successful demonstrations of applying quantum computation in various proof-of-principle experiments and numerical simulations~\cite{Martinez2016, Klco2018, Atas2022, Surace_2021,mildenberger2022probingconfinementmathbbz2lattice, Farrell_PhysRevD.107.054512}.
Simulating scattering processes for lattice field theories on a quantum computer poses a number of particular challenges.
Firstly, one has to be able to simulate the evolution of some initial state under the Hamiltonian describing the theory.
Secondly, the particles of the theory are typically highly nontrivial objects, and preparing the qubits on a digital quantum device in an initial state that corresponds to two particles with momenta, such that they propagate towards each other and scatter, is not an easy task. Reference~\cite{jordan_2019, Jordan_2018} first proposed a framework for scalar field scattering, employing trotterization and adiabatic time evolution to prepare the wave packet. However, this approach is resource-intensive and not feasible for current NISQ devices.

 In this paper we provide an efficient setup towards addressing the challenge of state preparation in the fermionic scattering using a staggered lattice discretization of the Thirring model~\cite{Thirring1958} as a test bed.
Using Givens rotation~\cite{PhysRevA.92.062318, Givens_rotation_1_PhysRevLett.120.110501, Jiang2018,2020}, we show how to prepare localized fermionic wave packets with momentum exactly on a digital quantum computer such that they propagate through a lattice during time evolution.
Starting from the ground state of the Hamiltonian, which can be obtained by the variational quantum eigensolver (VQE)~\cite{Peruzzo2014} or some other method, this allows us to simulate scattering between particles and antiparticles for fermionic models.
By performing numerical simulations of the quantum system on a classical computer, we demonstrate the elastic scattering of the fermions and we examine the entropy production during the evolution in the Thirring model.
Our approach demonstrates the Thirring model can straightforwardly be generalized to arbitrary fermionic models.

The paper is organized as follows.
In Sec.~\ref{sec:model}, we introduce the staggered discretization for the Thirring model that we are studying.
Subsequently, we show how to create Gaussian wave packets with appropriate momentum to make the two particles collide in Sec.~\ref{sec:scattering}.
Our numerical results demonstrating the procedure are shown in Sec.~\ref{sec:results}. Finally, we conclude in Sec.~\ref{sec:conclusion} and provide a future outlook.

\section{Model system and lattice formulation\label{sec:model}}

We consider the massive Thirring model in (1+1) dimensions.
The continuum Lagrangian $\mathcal{L}$ for this model reads~\cite{Thirring1958, Mari_Carmen}
\begin{equation}
    \mathcal{L} =i\overline{\psi}\gamma^{\mu}\partial_{\mu}\psi - m\overline{\psi}\psi - \frac{\lambda}{2}(\overline{\psi}\gamma_{\mu}\psi)(\overline{\psi}\gamma^{\mu}\psi),
    \label{eq:lagrangian_continuum}
\end{equation}
where $\psi$, $\overline{\psi} \equiv \psi^{\dagger}\gamma^0$ are two-component Dirac spinors, $m$ is the bare fermion mass, and $\lambda$ the dimensionless four-fermion coupling constant. The spinor components fulfill the usual anti-commutation relations for fermions,
\begin{align*}
    \{\psi_\alpha^\dagger(x),\psi_\beta(y)\} &= \delta_{\alpha,\beta}\delta(x-y),\\
    \{\psi_\alpha(x),\psi_\beta(y)\} &= 0,
\end{align*}
where $\alpha, \beta \in\{0,1\}$ label the components of the spinor. 
In (1+1) dimensions, the index $\mu$ takes values 0 and 1, and the matrices $\gamma^\mu$ correspond to two of the Pauli matrices $\{\sigma^x$, $\sigma^y$, $\sigma^z\}$ up to a phase factor.
Here we choose to work with the representation $\gamma^0 = \sigma^z$, $\gamma^1= i\sigma^y$. For a positive value of the coupling $g$, the model is known to exhibit an attractive force between fermions and antifermions~\cite{Coleman_1975}.

In order to address the theory numerically on a quantum computer, we need to adopt a Hamiltonian lattice formulation of the model.
Following Ref.~\cite{Mari_Carmen}, we adopt the Kogut-Susskind staggered formulation, where the components of the spinors are distributed to different lattice sites in order to avoid the doubling problem~\cite{Kogut1975, Susskind1977}. 
For a periodic lattice with $N$ sites and lattice spacing $a$, the lattice discretization of the Thirring model reads~\cite{Mari_Carmen},
\begin{equation}
    \begin{aligned}
    H &= \sum_{n=0}^{N-1}\left(\frac{i}{2a} \left( \xi_{n+1}^{\dagger}\xi_{n} - \xi_n^{\dagger}\xi_{n+1}\right) + (-1)^n m \ \xi_n^{\dagger}\xi_n \right)\\
    &+ \sum_{n=0}^{N-1} \frac{g(\lambda)}{a} \xi_n^{\dagger}\xi_n \xi_{n+1}^{\dagger} \xi_{n+1},    
    \end{aligned}
    \label{eq:hamiltonian_lattice_fermionic}
\end{equation}
where $g(\lambda) = \cos\bigl((\pi - \lambda)/2\bigr)$ with $-\pi<\lambda\leq\pi$ and $\xi_n$ is a single-component fermionic field on site $n$, and we identify $\xi_N^{\dagger} \equiv \xi_0^{\dagger}$. The fermionic creation and annihilation operators satisfy the anti-commutation relations,
\begin{equation}
\{\xi_n^{\dagger}, \xi_{n^{\prime}}\} = \delta_{n, n^{\prime}}, \ \  \{\xi_n^{\dagger}, \xi_{n^{\prime}}^{\dagger}\} = 0, \ \ \{\xi_n, \xi_{n^{\prime}}\} = 0.
\end{equation}

Note that for $g=0$ the four-fermion interaction term in Eq.~\eqref{eq:hamiltonian_lattice_fermionic} vanishes, and the resulting Hamiltonian corresponds to a staggered discretization of $N/2$ free Dirac fermions, which is quadratic in the fermion fields\footnote{Since we are working with a staggered discretization that separates the two components of the Dirac spinor to the odd and even sublattices, $N$ is considered to be even.}.
Thus, this case can be solved with free-fermion techniques both analytically~\cite{Lieb1961} or on a quantum device~\cite{Verstraete2009}.
Moreover, in this limit, the interpretation of the staggered lattice discretization in terms of particles and antiparticles becomes more transparent.
Occupied even sites contribute $m$ to the energy and, thus, can be interpreted as particles with mass $m$.
In contrast, occupied odd sites contribute $-m$ to the energy and correspond to the Dirac sea. Empty odd sites hence represent to a hole in the Dirac sea, i.e.,\ an antiparticle with mass $m$~\cite{Zohar2015a}.

As conventional qubit-based quantum hardware cannot directly address fermionic degrees of freedom, we map the degrees of freedom to spin operators using a Jordan-Wigner transformation,~\cite{jordan1993paulische}
\begin{equation}
    \begin{aligned}
        \xi_n^{\dagger} &= \prod_{l<n} \ \sigma_l^z \sigma_n^{-}, \quad
        \xi_n  &= \prod_{l<n} \ \sigma_l^z \sigma_n^{+}, 
    \end{aligned}
    \label{eq:jw}
\end{equation}
where $\sigma^\pm_l =\left(\sigma^x_l\pm i\sigma^y_l\right)/2$. Inserting the expression above into Eq.~\eqref{eq:hamiltonian_lattice_fermionic}, we obtain the Hamiltonian in terms of Pauli matrices
\begin{equation}
    \begin{aligned}
    	H &= \frac{i}{2a} \sum_{n=0}^{N-2} \left( \sigma_{n+1}^- \sigma_n^+ - \sigma_n^- \sigma_{n+1}^+ \right)  \\
    	   \,+& \frac{i}{2a} \left( \sigma_0^- \sigma_1^z \dots \sigma_{N-2}^z \sigma_{N-1}^+ - \sigma_{N-1}^- \sigma_{N-2}^z \cdots \sigma_{1}^z \sigma_{0}^+\right) \\
    	   \,+& \frac{m}{2} \sum_{n=0}^{N-1} (-1)^n \left( \mathds{1} - \sigma_n^z  \right) \\
          \,+& \frac{g}{4a}\sum_{n=0}^{N-1} \left( \mathds{1} - \sigma_n^z  \right) \left( \mathds{1} - \sigma_{n+1}^z  \right),
    \end{aligned}
    \label{eq:Hamiltonian_Paulis}
\end{equation}
where we again identify $\sigma_N^i = \sigma_0^i$, $i\in\{x,y,z\}$, and tensor products between operators are assumed. Note that the non-local Pauli-strings in the second line of Eq.~\eqref{eq:Hamiltonian_Paulis} arise due to the periodic boundary conditions, and are a consequence of the Jordan-Wigner transformation of the terms $\left( \xi_0^{\dagger}\xi_{N-1} - \xi_{N-1}^{\dagger}\xi_0 \right)$ wrapping around the boundary. For the rest of this paper, we consider without loss of generality, $a=1$.

\section{Simulating real-time dynamics of scattering on a digital quantum computer\label{sec:scattering}}

Simulating the out-of-equilibrium dynamics following a particle collision is a highly nontrivial task, which is to a large extent inaccessible with Monte Carlo (MC) methods due to the sign problem. Thus, there has been an ongoing effort to utilize methods from quantum information theory to simulate real-time evolution of scattering processes~\cite{Belyansky2023,Simone_prd} using the Hamiltonian formulation. In order to be able to simulate scattering processes, the following steps have to be accomplished.
\begin{enumerate}
    \item Preparing the ground state of the Hamiltonian, $\ket{\upOmega}$, which we refer to as the vacuum state.
    \item{Generating an initial state $\ket{\psi(0)} = D^{\dagger} C^{\dagger}  \ket{\upOmega}$ that corresponds to the vacuum state with two wave packets at arbitrary distance, which represent particles  created by $C^{\dagger}$ (antiparticles  created by $D^{\dagger}$)} with appropriately chosen momenta.
    \item{Performing a real-time evolution on the initial state $\ket{\psi(t)} = e^{-iHt}\ket{\psi(0)}$ such that the particles will propagate and interact with each other over time.}
    \item {Measuring relevant observables in the state $\ket{\psi(t)}$.}
\end{enumerate}
Mature approaches exist for performing step 1 on current quantum hardware, e.g.,\ by means of VQE~\cite{Peruzzo2014, Klco2018, Kokail2018,Angelides2023a,Farrell2023} or by applying variational imaginary time evolution~\cite{McArdle_2019, Gacon2023, Mazzola2021}. Similarly, the real-time simulation in step 3 can be simulated in general by trotterizing the evolution operator~\cite{Nielsen2002,Miessen_2023, Martinez2016, Klco2018, Nguyen2022, Farrell2024}, or by means of variational time evolution methods~\cite{Li2017,Barison2021,Miessen_2023}. Moreover, step 4 can also be done efficiently on a quantum computer for local observables.

Here we focus on some aspects concerning the steps 2--4, and assume the ground state can be prepared with some unitary $U_0$ from the state corresponding to all qubits initialized in state $\ket{0\dots0} \equiv \ket{0}$ such that $\ket{\upOmega} = U_0 \ket{0}$. Section~\ref{sec_prep} reports on the preparation of the initial state corresponding to two particles with opposite momenta. The procedure consists of two steps. First, we consider the free fermionic case corresponding to $g=0$. In this case, the model can be solved analytically, which allows us to construct exact operators creating a wave packet for particles and antiparticles. While these operators generate true (anti)particles only in the non-interacting case, they will also provide a valid approximation for a particle in the interacting case. 
In Section~\ref{sec_circuits} we then discuss the implementation of these states on a quantum device and provide the corresponding circuits using Givens rotations entangling gates. The implementation of the quantum dynamics circuits is described in Section~\ref{sec_dyn}, while Section~\ref{sec_td_obs} reports on the calculation of the time-dependent observables.

\subsection{Free Gaussian fermion and antifermion wave packets in the  staggered encoding\label{sec_prep}}

Let us first consider the non-interacting case $g=0$, for which Eq.~\eqref{eq:hamiltonian_lattice_fermionic} corresponds to the staggered discretization of $N/2$ free Dirac fermions. Our goal is to derive operators which create a localized particle or antiparticle in the form of a Gaussian wave packet with associated momentum, acting on the vacuum state. This can be easily done using the plane wave solutions of the theory in momentum space. Reference~\cite{Simone_prd} solved the model analytically, and in the rest of this subsection we will briefly review their results and show the construction of Gaussian wave packets.

Note that Eq.~\eqref{eq:hamiltonian_lattice_fermionic} is invariant under translations by two lattice sites due to the staggered mass term. Thus, it can be solved using standard methods by first performing a Fourier transformation followed by a Bogoliubov transformation. Doing this, we obtain two sets of creation operators, $c_k^\dagger$ and $d_k^\dagger$, in momentum space which correspond to the creation operators for particles and antiparticles with momenta $k\in  \frac{2\pi}{N}  \times \{ -\lfloor\frac{N}{4}\rfloor, \cdots \lceil\frac{N}{4}\rceil-1  \}$. The plane wave solutions can be used to construct operators creating Gaussian wave packets. Their representation in momentum space is given by
\begin{equation}
    \begin{aligned}
        C^{\dagger}(\phi^{c}) &= \sum_k \phi_k^c c_k^{\dagger}, \quad\quad D^{\dagger}(\phi^d) = \sum_k \phi_k^d d_k^{\dagger},
    \end{aligned}
    \label{eq:creation_operators_gaussian_particle}
\end{equation}
where the coefficients are given by a Gaussian distribution with a complex phase
\begin{equation}
    \phi_k^{c(d)} =  \frac{1}{\sqrt{\mathcal{N}_k^{c(d)}}} e^{-ik\mu_n^{c(d)}} e^{-(k-\mu_k^{c(d)})^2 / 4\sigma_k^2}.
    \label{eq:coeff_momentum}
\end{equation}
The normalization factor
\begin{equation}
    \mathcal{N}_k^{c(d)} = \sum_k \left |\phi_k^{c(d)} \right|^2
\end{equation}
ensures that $|\phi_k^{c(d)}|^2$ is a probability density and $\sum_k |\phi_k^{c(d)}|^2 = 1$. The operator $C^\dagger(\phi_c)$ $(D^\dagger(\phi_d)$) will create a Gaussian wave packet with a mean momentum $\mu_k^{c(d)}$ and width $\sigma_k$ in momentum space, located around $\mu_n^{c(d)}$ in real space.

The operators in momentum space $c_k^\dagger$, $d_k^\dagger$ are related to the original annihilation and creation operators in real space by
\begin{equation}
    \begin{aligned}
        c_k^{\dagger} &= \frac{1}{\sqrt{N}} \sqrt{\frac{m+w_k}{w_k}} \sum_n e^{ikn} \left( \Uppi_{n0} + v_k \Uppi_{n1}\right) \xi_n^{\dagger}, \\
        d_k^{\dagger} &= \frac{1}{\sqrt{N}} \sqrt{\frac{m+w_k}{w_k}} \sum_n e^{ikn} \left( \Uppi_{n1} + v_k \Uppi_{n0}\right) \xi_n,
    \end{aligned}
    \label{eq:operator_transformation}
\end{equation}
where
\begin{equation} 
   v_k = \frac{\sin(k)}{m+w_k},\ \  w_k = \sqrt{m^2 + \sin^2(k)},
   \label{eq:wk_vk}
\end{equation}
and $\Uppi_{nl}$ is a projection operator defined as
\begin{align} 
    \Uppi_{nl} = \left(1+\left(-1\right)^{n+l}\right)/2,\quad\quad l\in\{0,1\}.
    \label{eq:projector}
\end{align}
This allows for obtaining an explicit representation of the operators creating a wave packet in position space, 
\begin{align} 
    C^{\dagger}(\phi^{c}) = \sum_n \tilde{\phi}_n^c \xi^{\dagger}_n, \quad\quad D^{\dagger}(\phi^d) = \sum_n \tilde{\phi}_n^d \xi_n,
    \label{eq:wave_packet_real_space}
\end{align}
where the normalized coefficients are given by
\begin{equation}
    \begin{aligned}
        \tilde{\phi}_n^c &= \frac{1}{\sqrt{N}} \sum_k \phi_k^c  \sqrt{\frac{m+w_k}{w_k}} e^{ikn} \left( \Uppi_{n0} + v_k \Uppi_{n1}\right), \\
        \tilde{\phi}_n^d &= \frac{1}{\sqrt{N}} \sum_k \phi_k^d  \sqrt{\frac{m+w_k}{w_k}} e^{ikn} \left( \Uppi_{n1} + v_k \Uppi_{n0}\right).
    \end{aligned}
    \label{eq:coeff_real_sapce}
\end{equation}
Looking at Eq.~\eqref{eq:wave_packet_real_space}, we see that $C^\dagger(\phi_c)$ ($D^\dagger(\phi_c)$) corresponds to a superposition of creation (annihilation) operators, thus showing that it creates a wave packet for particles (antiparticles). Note that the coefficients in Eq.~\eqref{eq:coeff_real_sapce} are again normalized,  $\sum_n |\tilde{\phi}_n^{c(d)}|^2 = 1$, hence $|\tilde{\phi}_n^{c(d)}|^2$ corresponds to a probability density.

\begin{figure*}
    \centering
    \includegraphics[width = 0.9\textwidth]{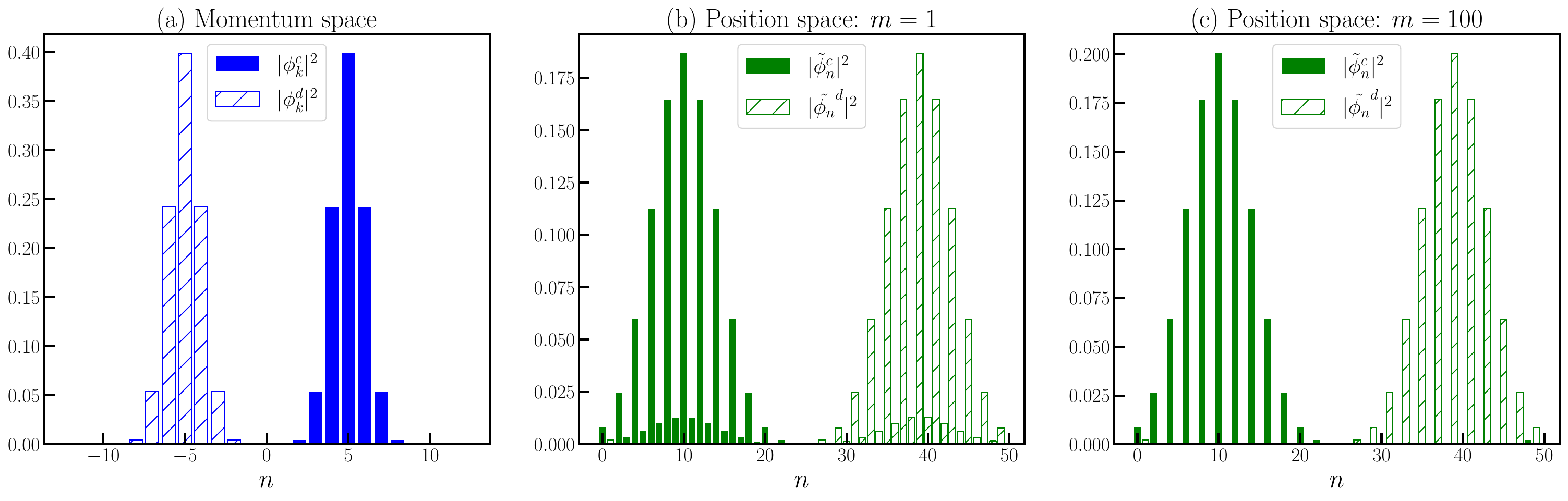}
    \caption{Probability density for Gaussian wave packets in momentum and position space for $N = 50$ sites. The solid histogram represents a Gaussian fermionic wave packet with $\mu_k^c = 5\times 2\pi/N$ and $\mu_n^c = 10$, while the dashed histogram corresponds a Gaussian antifermion wave packet with $\mu_k^d = -5\times 2\pi/N$ and $\mu_n^d = 39$, both with a width of $\sigma_k = 2\pi/N$. Panel (a) displays the probability density in momentum space, and panels (b), (c) show the probability density in position space for masses $m = 1$ and $100$. These panels illustrate that with increasing mass, resulting in a value of $v_k$ closer to zero, the probabiltiy distribution for the fermion (antifermion) in position space is increasingly suppressed on odd (even) sites.}
    \label{fig:Gaussian_wave_packet}
\end{figure*}

Focusing on the limit of infinite mass, $m \rightarrow \infty$, Eq.~\eqref{eq:wk_vk} reveals that for this case $v_k \rightarrow 0$. Together with the properties of the projectors in Eq.~\eqref{eq:projector}, we see that in this limit $C^\dagger(\phi^c)$ ($D^\dagger(\phi^d)$) only affects the even (odd) sites. Thus, in this limit the particles are only located at the even sites whereas the antiparticles reside on the odd sites. For a finite mass one finds $0 < |v_k| \leq 1$, hence the fermion and antifermion are located on both even and odd sites, with the amplitude of fermion (antifermion) being by suppressed the factor $|v_k|$ on odd (even) sites. Figure~\ref{fig:Gaussian_wave_packet} illustrates this effect by showing the probability density $|\phi_k^{c(d)}|^2$ in momentum and $|\tilde{\phi}_n^{c(d)}|^2$ in position space for two values of the mass.

A state corresponding to the vacuum with a fermion and antifermion wave packet superimposed on top can now be obtained by acting with the operators $C(\phi^c)$ and $D(\phi^d)$ on the state $\ket{\upOmega}$ with,
\begin{align}
    \ket{\psi(0)} = D^{\dagger}(\phi^d) \ C^{\dagger}(\phi^c) \ket{\upOmega}.
\end{align}
In order to prepare this state on a quantum device, one has to be able to implement the action of the operators $C^{\dagger}(\phi^c)$ and $D^{\dagger}(\phi^d)$ on a given fermionic state. In the following we discuss how this can be achieved with Givens rotations~\cite{PhysRevA.92.062318,Givens_rotation_1_PhysRevLett.120.110501,Jiang2018}.

\subsection{Givens rotation for preparing fermionic states}
\label{sec_circuits}

One of the most efficient ways to implement the action of linear combinations of the fermionic operators on a fermionic state is to use Givens rotations~\cite{PhysRevA.92.062318}. This approach allows for the realization of the resulting fermionic states with a quantum circuit of linear depth in the system size, $N$~\cite{Givens_rotation_1_PhysRevLett.120.110501, Jiang2018}.

The linear combination of fermionic operators in Eq.~\eqref{eq:wave_packet_real_space} can be interpreted as a unitary transformation of the original creation and annihilation operators
\begin{align}
    V(u)\xi_0^\dagger V^\dagger(u) = \sum_n \xi^\dagger_n u_{n0}
    \label{eq:unitary_transformation_xi}
\end{align}
where 
\begin{align}\label{eq: V_u}
    V(u) = \exp\left(\sum_{nl}\xi^\dagger_n \left[\log u\right]_{nl}\xi_l\right)
\end{align}
with a unitary matrix $u\in\mathbb{C}^{N\times N}$. Choosing the first column of $u$ to be equal $\tilde{\phi}_n^c$ ($\tilde{\phi}_n^d$), Eq.~\eqref{eq:unitary_transformation_xi} corresponds exactly to the creation operators for Gaussian wave packets for fermion (antifermions) in Eq.~\eqref{eq:wave_packet_real_space}. Note that this is always possible since $\sum_n|\tilde{\phi}_n^{c(d)}|^2 = 1$ and thus the $\tilde{\phi}_n^{c(d)}$ can be interpreted as the entries of a complex unit vector in $\mathbb{C}^N$. Furthermore, since we do not care about the other columns of $u$, we can simply choose these such that the columns of $u$ form an orthonormal basis of $\mathbb{C}^N$.

The map $V(u)$ is a homomorphism under matrix multiplication, i.e.,\ $V(u \cdot u^{\prime}) = V(u) \vdot V(u^{\prime})$. Thus, in order to obtain a decomposition in $V(u)$ in unitary gates that can be implemented on a quantum computer, we can decompose the matrix $u$ in a product of unitaries acting nontrivially only on a few sites. This can be done with Givens rotations~\cite{Givens_rotation_1_PhysRevLett.120.110501, Jiang2018}, as detailed in Appendix~\ref{app:Givens_rotation}. Since we care only about the first column of $u$, corresponding to $\tilde{\phi}_n^c$, we hereafter will write $ V(\tilde{\phi}_n^c)$ instead of $V(u)$.

In summary, we can decompose the matrix $u$ into a product of a diagonal matrix $p(\vec{\beta})$ and Givens rotation matrices, $r_n(\theta_n)$, $n=1,\dots,N-1$. This allows us to construct a decomposition of the operator $V(u)$ using the fact that it is a homomorphism under matrix multiplication. In particular, the structure of the Givens rotation matrices is such that $V(r_n(\theta_n))$ acts non-trivially only on two neighboring sites. 
More specifically, if we write the complex Gaussian amplitudes in position space in polar form, $\tilde{\phi}_n^c = a_ne^{-i\beta_n}$, $n=0,\dots, N-1$, with absolute value $a_n$ and phase angle $\beta_n$, we can decompose the operator $V$ as,
\begin{align}\label{eq: decomp_Vc}
    V(\tilde{\phi}_n^c) &= V^{\dagger}(p(\vec{\beta}))  V^{\dagger}(r_{N-1}) \cdots V^{\dagger}(r_1).
\end{align}
In the expression above $\vec{\beta}$ is a real vector whose entries are the phase angles of the Gaussian amplitudes and the terms correspond to
\begin{align}\label{eq: Vp_Vr}
    V^{\dagger}(p(\vec{\beta})) & = \exp\left( -i\sum_n \beta_n\  \xi_n^{\dagger} \xi_n\right), \\
    V^{\dagger}(r_n) &= \exp\left( \theta_n  [\xi^{\dagger}_{n-1} \xi_n - \xi^{\dagger}_n \xi_{n-1}] \right),
\end{align}
where $\theta_n$ is the Givens rotation angle given by $\theta_n = \arctan{\left( -a_n/a_{n-1} \right)}$. 

\begin{figure}[htp!]
    \centering
    \includegraphics[width=0.98\columnwidth]{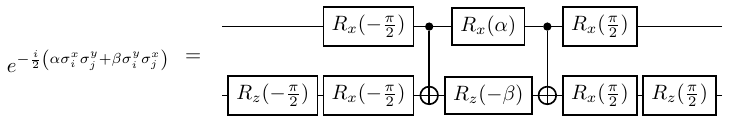}
    \caption{Decomposition of a gate of the form $\exp\left(-i\left( \alpha \sigma^x_i \sigma^y_j +  \beta \sigma^y_i \sigma^x_j\right)/2\right)$ into Pauli rotation and CNOT gates.}
    \label{fig:givens}
\end{figure}

Using the Jordan-Wigner transformation from Eq.~\eqref{eq:jw}, the operator $V^{\dagger}(p(\vec{\beta}))$ can be mapped to a set of single-qubit rotation gates (up to a global phase)
\begin{align}\label{eq: Vp_Rz}
    V^{\dagger}(p(\vec{\beta})) = \exp\left( i\sum_n \frac{\beta_n}{2}\  \sigma_n^z\right)=\prod_nR_z(-\beta_n),
\end{align}
where $R_z(\theta) = \exp(-i \theta \sigma^z/2)$ is the standard Pauli rotation gate about the $z$-axis. For $V^{\dagger}(r_n)$ the Jordan-Wigner transformation yields a particle-number conserving two-qubit gate
\begin{align}
   V^{\dagger}(r_n(\theta_n)) = \exp\left( i \frac{\theta_n}{2}  \left(\sigma_{n-1}^x\sigma_n^y - \sigma_{n-1}^y\sigma_n^x \right) \right).
   \label{eq:two_qubit_gate}
\end{align} 
The two-qubit gate in Eq.~\eqref{eq:two_qubit_gate} above can be easily expressed by single-qubit Pauli rotation gates, $R_P(\theta)=\exp(-i P\theta/2)$, $P\in\{\sigma^x,\sigma^y, \sigma^z\}$, and CNOT gates as shown in Fig.~\ref{fig:givens}. 
Analogously, we can obtain an expression for $V(\tilde{\phi}_n^d)$ and also decompose it into a quantum circuit using Givens rotations.

Finally, we can rewrite the operators creating a Gaussian fermion (antifermion) wave packet as
\begin{align}
    \begin{aligned}
        C^{\dagger}(\phi^c) & = \sum_{n=0}^{N-1} \xi_n^{\dagger} \tilde{\phi}_n^c  = V(\tilde{\phi}_n^c) \sigma_0^- V^{\dagger}(\tilde{\phi}_n^c), \\
        D^{\dagger}(\phi^d) & = \sum_{n=0}^{N-1} \xi_n \tilde{\phi}_n^d  = V(\tilde{\phi}_n^{d\ast}) \sigma_0^+ V^{\dagger}(\tilde{\phi}_n^{d\ast}),
    \end{aligned}
    \label{eq: givens_fermion_op}
\end{align}
where $\tilde{\phi}_n^{d\ast}$ is the complex conjugate of $\tilde{\phi}_n^{d}$, and we translated the fermionic operators from Eq.~\eqref{eq:unitary_transformation_xi} to spin operators,  again using the Jordan-Wigner transformation. The unitary operators $V$ on the right-hand side can be implemented as quantum circuit with a combination of Givens rotations~\cite{PhysRevA.92.062318, Givens_rotation_1_PhysRevLett.120.110501, Jiang2018}, as outlined above. Note that the Pauli raising operator $ \sigma_0^+$ and its Hermitian conjugate are non-unitary and therefore cannot be directly implemented on a quantum device. Nevertheless, the form of the operators $C^{\dagger}(\phi^c)$ and $D^{\dagger}(\phi^d)$ we derived here is advantageous to obtain the time-dependent expectation values observables on a quantum device, as discussed in Sec.~\ref{sec_td_obs}.

There is an orthogonality between the coefficients of fermion and antifermion wave packets, illustrated by $\sum_n \tilde{\phi}_n^d \cdot \tilde{\phi}_n^c = 0$, which is proved in Appendix~\ref{app: orthogonality}. This property allows for a reduction in circuit depth when preparing one fermion and one antifermion compared to the approach in Eq.~\eqref{eq: givens_fermion_op}. As detailed in Appendix~\ref{app: decompose_Ut_Givens}, a simplified circuit approach was employed for wave packet preparation in our hardware experiments for the fermion-antifermion scattering. However, the coefficients between two fermions or two antifermions are not orthogonal, so we still use the formula in Eq.~\eqref{eq: givens_fermion_op} for a comprehensive explanation of our methodology.

\subsection{State initialization and simulation of dynamics\label{sec_dyn}}

In order to simulate the dynamics of scattering processes, the initial state we want to generate is given by $\ket{\psi(0)} = D^{\dagger}(\tilde{\phi}_n^d) C^{\dagger}(\tilde{\phi}_n^c)\ket{\upOmega}$. For the free fermionic case discussed in Sec.~\ref{sec_prep}, which corresponds to the Thirring model with coupling $g=0$, this state exactly represents the ground state with two Gaussian wave packets -- one corresponding to an antifermion and another corresponding to a fermion -- superimposed on top. For the case of non-vanishing coupling, applying the operators $D^{\dagger}(\tilde{\phi}_n^d)$ and $C^{\dagger}(\tilde{\phi}_n^c)$ to the ground state will no longer produce exactly a state with a particle-antiparticle wave packet.
For the weak coupling regime, $|g/m| \ll 1$, we still expect to obtain a good approximation for the true particle-antiparticle wave packet. Similar observations have been made for the Schwinger model in Ref.~\cite{Papaefstathiou2024}.

In the strong coupling regime, $|g/m| \gg 1$, applying the operators $D^{\dagger}(\tilde{\phi}_n^d)$ and $C^{\dagger}(\tilde{\phi}_n^c)$ will likely not provide stable (anti-)particle wave packets. For the rest of the paper, we choose $m$ and $g$ to ensure that we are in a regime where we expect our approach to be stable particle-antiparticle wave packets representing a good approximation for physical fermions antifermions.
This will be verified post facto in our numerical simulations in Sec.~\ref{sec:results}.

Here we focus on the description of the dynamics of such an initial state, which is given by
\begin{equation}
    \ket{\psi(t)} = e^{-iHt} \ket{\psi(0)} = e^{-iHt} D^{\dagger}(\tilde{\phi}_n^d) C^{\dagger}(\tilde{\phi}_n^c)\ket{\upOmega}.
    \label{eq:time_evolved_state}
\end{equation}
As mentioned before, the dynamics can be implemented via standard algorithms, such as Trotter decomposition or variational time evolution. We will not discuss these approaches in detail, and estimating the resources required for these methods as well as assessing their performance for the general case is left for future work.

For Hamiltonians that are at most quadratic in the fermionic operators, the time evolution can be implemented more efficiently. On the one hand, the considered Hamiltonian is diagonal in momentum space. Hence, if the evolution is computed in momentum space, $\exp(-iHt)$ simply corresponds to a phase factor. On the other hand, in position space, $\exp(-iHt)$ has the form of Eq.~\eqref{eq: V_u}. Thus, it can be decomposed using Givens rotations as demonstrated for the operators creating wave packets in the previous section. The resulting circuit implements the time evolution operator for arbitrary $t$ without any approximation at a constant circuit depth $\mathcal{O}(4N)$ using $\mathcal{O}(N^2)$ CNOT gates. Thus, for the non-interacting case this approach presents a more promising avenue towards simulating the dynamics on quantum hardware than using methods based on Trotter decomposition or variational time evolution. The technical details for implementing the time evolution operator in the non-interacting case using Givens rotation are presented in Appendix~\ref{app: decompose_Ut_Givens}.

In general, we find the time-evolved state by combining Eqs.~\eqref{eq: givens_fermion_op} and \eqref{eq:time_evolved_state},
\begin{align}
    \begin{aligned}
    \ket{\psi(t)} &= U_3 \sigma_0^+  U_2 \sigma_0^- U_1 \ket{0}\\
                  &= \frac{1}{4} U_3 (\sigma_0^x + i \sigma_0^y)  U_2 (\sigma_0^x - i \sigma_0^y) U_1 \ket{0},
    \end{aligned}
    \label{eq:psi_evolved}
\end{align}
where we have defined the unitaries
\begin{align}
    U_3 &= e^{-iHt} V(\tilde{\phi}_n^{d\ast}), \\
    U_2 &= V^{\dagger}(\tilde{\phi}_n^{d\ast})V(\tilde{\phi}_n^c), \\
    U_1 &=  V^{\dagger}(\tilde{\phi}_n^c) U_0,
\end{align}
and used the fact that the ground state can be prepared via some unitary $U_0$ from the state $\ket{0}$. Note that the state $\ket{\psi(t)}$ in Eq.~\eqref{eq:psi_evolved} is not a normalized state that can be prepared on a quantum device, as the operators $\sigma_0^\pm = (\sigma_0^x \pm i \sigma_0^y)/2$ are not unitary. Nevertheless, it is possible to obtain expected values of observables efficiently from a quantum device, as we discuss in the following section.

\subsection{Evaluation of time-dependent observables\label{sec_td_obs}}

In general, we want to calculate the expectation value of some observable $O$ in the state $\ket{\psi(t)}$. Using Eq.~\eqref{eq:psi_evolved}, the expectation value of $O$ can be expressed as
\begin{align}
    \begin{aligned}
    \langle O \rangle &= \frac{1}{\sqrt{\mathcal{N}}}\bra{\psi(t)} O \ket{\psi(t)}  \\
    &= \frac{1}{\sqrt{\mathcal{N}}}\times \frac{1}{16}\sum_{\bar \mu}  c_{\bar \mu} \langle O \rangle_{\bar \mu},
    \end{aligned}
    \label{eq: expectation}
\end{align}
where we have explicitly taken into account the normalization factor  $\mathcal{N} = |\bra{\psi(t)} \ket{\psi(t)}|^2$ since $\ket{\psi(t)}$ is not normalized. The symbol $\bar \mu$ corresponds to a multi-index $\bar \mu= (\mu_1, \mu_2, \mu_3, \mu_4)$ with $\mu_\ell \in \{x, y\}$, and $\langle O \rangle_{\bar \mu}$ is a short-hand notation for
\begin{align}
    \langle O \rangle_{\bar \mu} &= \bra{0} (U_1^\dagger\sigma_0^{\mu_1} U_2^{\dagger} \sigma_0^{\mu_2}  U_3^{\dagger})\, O\, (U_3  \sigma_0^{\mu_3}  U_2 \sigma_0^{\mu_4}U_1) \ket{0}.
    \label{eq:expectation_summand}
\end{align}
The coefficients in Eq.~\eqref{eq: expectation} are given by $c_{\bar \mu} = i^{\delta_{\mu_1,y}}(-i)^{\delta_{\mu_2,y}}i^{\delta_{\mu_3,y}}(-i)^{\delta_{\mu_4,y}}$, where $\delta_{\mu_\ell,y}$ is nonzero iff $\mu_\ell=y$, and evaluate to $\{\pm 1, \pm i\}$, depending on the combination of Pauli matrices involved. Note that in Eq.~\eqref{eq:expectation_summand} all operators acting on the state $\ket{0}$ are unitary and can be implemented a quantum computer. 

For the case $\mu_1 = \mu_4$ and $\mu_2 = \mu_3$, Eq.~\eqref{eq: expectation} represents nothing but the expectation value of observable $O$ in the state $\ket{\chi} =U_3  \sigma_0^{\mu_2}  U_2 \sigma_0^{\mu_1}U_1 \ket{0}$, which can be directly measured on a quantum device.

\begin{figure}
    \centering
    \includegraphics[width=0.98\columnwidth]{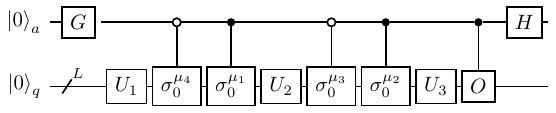}
    \caption{Circuit for measuring the terms in Eq.~\eqref{eq: expectation} utilizing an ancilla qubit $a$. The expected value of $\left( c_{\bar \mu} \langle O \rangle_{\bar \mu} + \text{h.c.} \right)$ can be obtained by measuring the expected value of $\sigma^z$ on the ancilla qubit. The gate $G$ corresponds to the Hadamard gate $H$ if $c_{\bar \mu} \in \{\pm 1\}$, and to $G \equiv R_x(\pi/2)$ if $c_{\bar \mu} \in \{\pm i\}$.}
    \label{fig: Hadamard test circuit}
\end{figure}

In all other cases, we can find the expected value of the non-Hermitian operator with the following observation. Since we sum over all possible combinations in Eq.~\eqref{eq:expectation_summand} and the Pauli matrices are self-adjoint, the sum is going to contain the Hermitian conjugate of each summand. Hence, we do not need to determine the individual summands; it is sufficient to obtain $(c_{\bar \mu}\langle O \rangle_{\bar \mu} + \text{h.c.})$ from the quantum device. This can be done with a circuit similar to that used for the Hadamard test~\cite{Chiesa_2019,Endo2019}. The corresponding quantum circuit requires the addition of an ancilla qubit, which is entangled with the qubits encoding the state wave function, as shown in Fig.~\ref{fig: Hadamard test circuit}. A subsequent measurement of the expectation value of the Pauli operator $\sigma^z$ yields the desired quantity. The normalization factor $\mathcal{N}=\bra{\psi(t)} \ket{\psi(t)}$ can be computed in a similar manner by simply setting observable $O=\mathds{1}$.

We now specialize our investigation on the case of free staggered fermions, corresponding to $g=0$ in Eq.~\eqref{eq:hamiltonian_lattice_fermionic}. The resulting ground state is trivial in momentum space and is simply given by the state with no (anti)fermionic excitations; see Eq.~\eqref{eq: annihilation_vaccum} in Appendix~\ref{app:momentum_and_fock_space} for details. Hence, annihilating a (anti)fermion in this state is not possible, and acting with the operators $C$, $D$ on the vacuum state will result in zero. This allows us to rewrite the initial state as,
\begin{align}
    \begin{aligned}
    &\ket{\psi(0)} = D^{\dagger}(\tilde{\phi}_n^d) C^{\dagger}(\tilde{\phi}_n^c) \ket{\upOmega} \\
    &= \left(D^{\dagger}(\tilde{\phi}_n^d) + D(\tilde{\phi}_n^d)\right) \cdot
       \left(C^{\dagger}(\tilde{\phi}_n^c) + C(\tilde{\phi}_n^c) \right) \ket{\upOmega} \\
    &= \left(V(\tilde{\phi}_n^{d\ast}) \sigma_0^x V^{\dagger}(\tilde{\phi}_n^{d\ast})\right) \cdot \left(V(\tilde{\phi}_n^c) \sigma_0^x V^{\dagger}(\tilde{\phi}_n^c)\right) \ket{\upOmega},
  \end{aligned}
  \label{eq:psi_init_free_case}
\end{align}
where in the third line we have substituted in Eq.~\eqref{eq: givens_fermion_op} with $\sigma_0^+ + \sigma_0^- = \sigma_0^x$. Notice that in the last line of Eq.~\eqref{eq:psi_init_free_case} all operators acting on the ground state are unitary. Since the time evolution operator is unitary as well, expectation values of observables $O$ in the time evolved state $\ket{\psi(t)} = e^{-iHt} \ket{\psi(0)}$ can simply be measured on a quantum device in a standard manner.

The simplification of the circuit discussed above applies only to the case of the free theory in case the initial state corresponds to a wave packet of one fermion and one antifermion. For more general settings considering wave packets for two or more fermions and antifermions, or when dealing with the interacting theory with a nontrivial ground state, the Hadamard test is a viable alternative for determining expectation values.

\section{Results\label{sec:results}}
In the following, we demonstrate the methods discussed above for the lattice Thirring model discussed in Sec.~\ref{sec:model} and whose Hamiltonian is given in Eq.~\eqref{eq:Hamiltonian_Paulis}. We will examine two regimes: first we focus on the case of $g=0$, for which the Hamiltonian corresponds to a staggered discretization of free Dirac fermions and is exactly solvable. Second, we investigate the interacting theory for $g\neq 0$, a regime where the solution of the model can no longer be obtained with analytical techniques. 
Lastly, we demonstrate the feasibility of our approach on current quantum hardware by simulating the evolution of the noninteracting case on IBM's quantum devices.
In both cases, we will consider initial states corresponding to the ground state with one fermion and one antifermion wave packet with opposite momenta imposed on top. Subsequently, we evolve those in time to observe the scattering of the two wave packets.

For the continuum Thirring model, the spectrum and the $s$-matrix can be computed analytically~\cite{TM_spectrum_A.Luther_1976, TM_S_matrix_Korepin_1979, TM_repulsive_korepin_1980}; see Refs.~\cite{Luther_A:1976, Luscher_M:1976pb} for slightly different interpretation.

The continuum Thirring model is equivalent to the quantum sine-Gordon model in the sector of zero topological charge \cite{Coleman1976, Mandelstam_1975}. Due to the infinite number of conservation laws in the sine-Gordon model, there can be no particle production during the scattering process~\cite{Zamolodchikov:1977, Karowski:1977}, and the sets of momentum of incoming and outgoing fermions are equal, which also hold for the lattice Thirring model~\cite{Luscher_M:1976pb}. Therefore, the particles can thus only exchange momentum during the scattering process. As such, we will restrict our discussion to the sector with constant particle number $\sum_n\xi^\dagger_n\xi_n = N/2$ which, after the Jordan-Wigner transformation, corresponds to $\sum_n \sigma_n^z = 0$.

To characterize the scattering process, we calculate several observables at each time step of the evolution. First, we will measure the site-resolved particle density $\langle \xi_n^{\dagger} \xi_n \rangle_t$, which in spin language is simply given by $\langle (\mathds{1}- \sigma^z_n)/2 \rangle_t$. In order to highlight the effect of wave packets compared to the ground state of the theory itself, we subtract the particle density of the ground state and consider the quantity 
\begin{align}
    \Delta \langle\xi_n^\dagger\xi_n\rangle_t = \bra{\psi(t)}\xi_n^\dagger\xi_n\ket{\psi(t)} - \bra{\upOmega}\xi_n^\dagger\xi_n\ket{\upOmega}.
    \label{eq:subtracted_fermion_density}
\end{align}
Second, we calculate at each step of the evolution the von Neumann entropy $S(n, t) = - \tr\left [ \rho_n(t) \log_2\rho_n(t) \right ]$ obtained for the reduced density operator $\rho_n$ describing the subsystem of the first $n$ qubits. This quantity provides a measure for entanglement between two subsystem $\mathcal{L} = \{l < n\}$ and $\mathcal{R} = \{l \geq n\}$.
To again highlight the effect caused by the addition of the wave packets, we subtract the entropy of the ground state $S_\upOmega(n)$ and consider,
\begin{align}
    \Delta S_1(n, t) = S(n,t) - S_\upOmega(n).
    \label{eq:subtracted_entropy}
\end{align}
In addition, to quantify the effect of the interaction of the wave packets, we consider the difference in von Neumann entropy obtained for a system with two wave packets compared to the sum of the entropy from two separate systems, each having either particle or an antiparticle but not both. The latter can be obtained by considering the initial states $\ket{\psi_C(0)} = C^\dagger(\phi^c)\ket{\upOmega}$ and $\ket{\psi_D(0)} = D^\dagger(\phi^c)\ket{\upOmega}$ with the same momenta and width of the wave packet as for the combined case and evolving those in time. The difference in entropy is then given by
\begin{align}
    \begin{aligned}
        \Delta S_2(n, t) &= \Delta S_1(n, t) \\
        &- \bigl(\Delta S_{1,C}(n,t) + \Delta S_{1,D}(n,t)\bigr),
    \end{aligned}
    \label{eq:subtracted_entropy_two_systems}
\end{align}
where $S_{1,(C,D)}(n,t)$ is the entropy for the time evolved state $\ket{\psi_{(C,D)}(0)}$, according to Eq.~\eqref{eq:subtracted_entropy}.

\subsection{Vanishing coupling: the free fermionic case}

Let us first consider the case of free staggered Dirac fermions. While the dynamics of the system in this regime can be simulated efficiently in momentum space using free fermionic techniques (see Appendix~\ref{app:momentum_and_fock_space}), we choose to work in position space to benchmark the techniques developed earlier. 

\begin{figure}[htp!]
    \centering
    \includegraphics[width=0.5\textwidth]{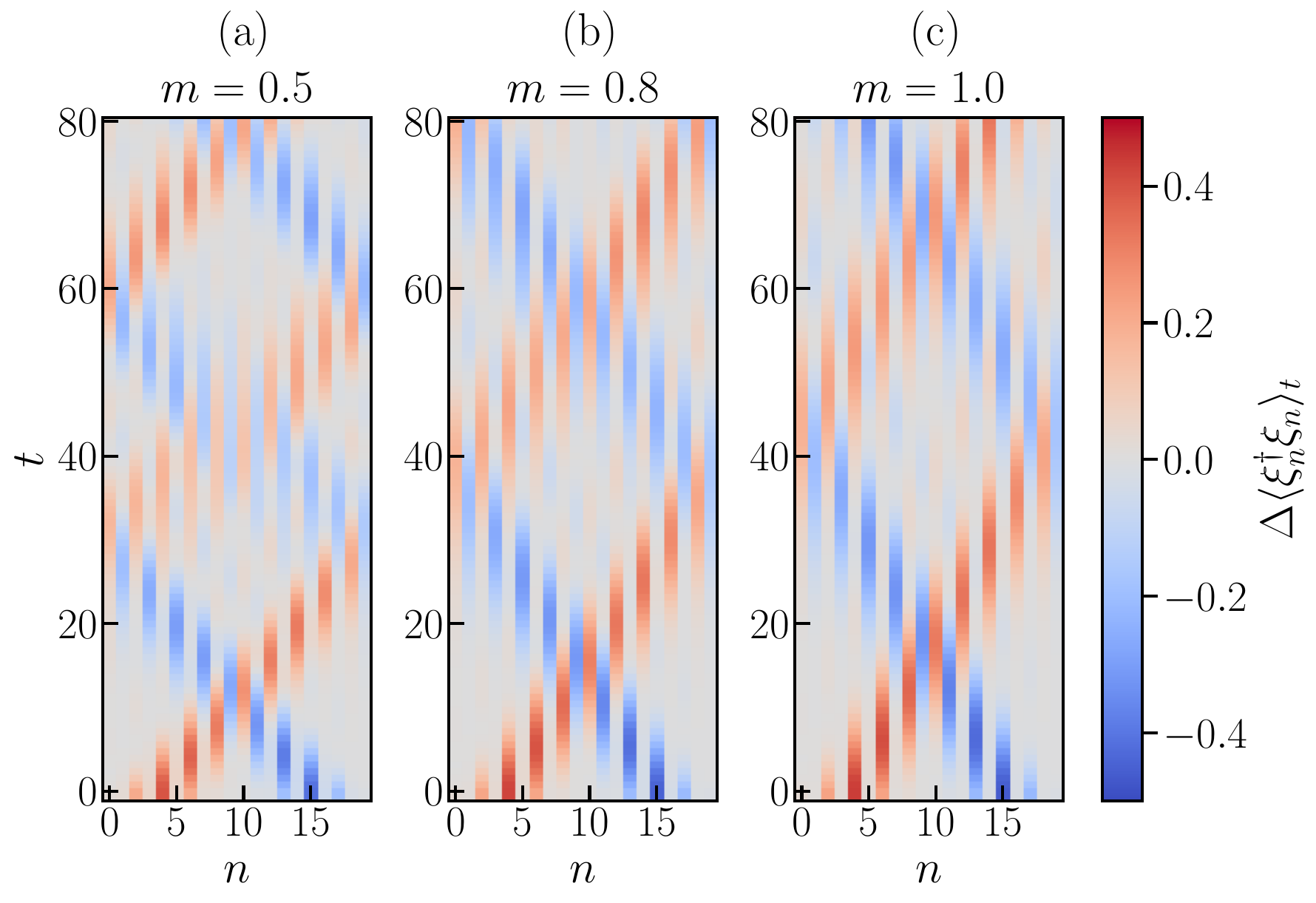}
    \caption{Site-resolved particle density as a function of time for the noninteracting case $g=0$ and number of lattice sites $N=20$. The different panels correspond to different masses (a) $m = 0.5$, (b) $0.8$, and (c) $1.0$. For all simulations, the initial Gaussian fermion wave packet is located around $\mu_n^c = 4$ with mean momentum $\mu_k^c = 2\times 2\pi/N $, and the initial Gaussian antifermion wave packet is located around $\mu_n^d = 15$ with mean momentum $\mu_k^d = -2\times 2\pi/N$. Both the fermion and antifermion wave packets have a width of $\sigma_k = 2\pi/N$ in momentum space.}
    \label{fig:charge_density_free_fermion}
\end{figure}

In order to demonstrate that the operators in Eq.~\eqref{eq: givens_fermion_op} successfully create a wave packet, we first simulate the evolution in time exactly. To this end, we obtain the ground state $\ket{\upOmega}$ of the Hamiltonian using exact diagonalization and subsequently compute the system's evolution via Eq.~\eqref{eq:time_evolved_state} by approximating the time evolution operator via a Taylor expansion up to second order, $\exp(-iH\Delta t) \approx 1- i\Delta t H -  H^2 (\Delta t)^2/2$. Here we use a time step of $\Delta t = 2\times 10^{-3}$ which is determined to be small enough to avoid noticeable numerical errors for the time scales we simulate. This allows us to easily reach a system size of 20 qubits without heavy memory requirement for the matrix representation of $e^{-iHt}$, the results for which for the site resolved particle density throughout the evolution for various masses are shown in Fig.~\ref{fig:charge_density_free_fermion}. 
The particle density clearly shows the presence of the two wave packets where the fermion (antifermion) manifests itself in an excess (lack) of particle density compared to the ground state. Moreover, we see the desired effect that the wave packets are moving towards each other, where the dynamics progress faster for smaller values of the mass, as expected. Since the fermions are noninteracting, the wave packets simply pass through each other before approaching one boundary and returning on the opposite due to periodic boundary conditions.
\begin{figure*}[htp!]
    \centering
    \includegraphics[width=0.8\textwidth]{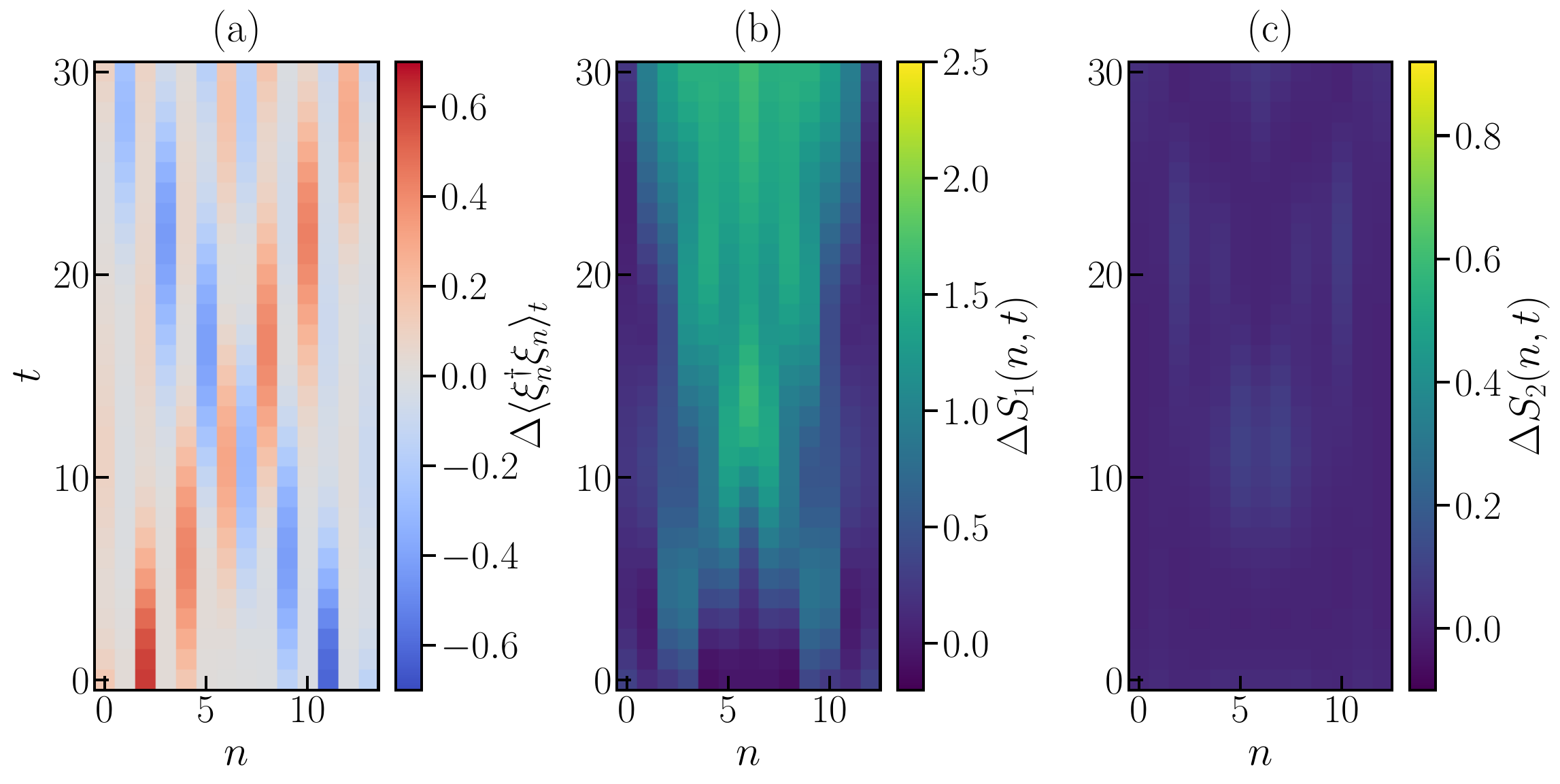}
    \caption{(a) Particle density, (b) entropy $\Delta S_1(n,t)$  and (c) subtracted entropy $\Delta S_2(n,t)$  for the non-interacting system with $14$ sites and $m=0.8$. The initial Gaussian fermion (antifermion) wave packet is located around the site $n=2$ ($n=11$) with the mean momentum $\mu_k^c = 2\pi/ N$  ($\mu_k^d = -2\pi/ N$). Both wave packets have the same width in momentum space $\sigma_k = 2\pi/N$.}
    \label{fig:free_entropy_14}
\end{figure*}

In order to gain more insight into the scattering process, we also consider the von Neumann entropy. Since computing the entropy in our direct approach requires constructing the reduced density operator, which in general is a dense matrix, we use a slightly smaller system size $N=14$ to reduce the memory overhead. Moreover, we do not perform a direct simulation with the matrix representation of the time evolution operator $e^{-iHt}$, but we simulate the circuit allowing us to compute the evolution exactly as discussed in Sec.~\ref{sec_dyn} and outlined in detail in Appendix~\ref{app: decompose_Ut_Givens}. For these simulations, we use state vector simulation in Qiskit~\cite{Qiskit} assuming an ideal quantum computer without shot noise. The results for the subtracted particle density and the von Neumann entropy are shown in Fig.~\ref{fig:free_entropy_14}. Despite the reduced system size, the particle density reveals that we initially still have two clearly separated wave packets which propagate towards each other during the evolution and eventually pass through one other. Inspection of the site-resolved entropy, $\Delta S_1(n,t)$, in Fig.~\ref{fig:free_entropy_14}(b), there is no entanglement between the wave packets at the beginning of the evolution but as they approach one another, they also become entangled. However, looking at the subtracted von Neumann entropy, $\Delta S_2(n,t)$, in Fig.~\ref{fig:free_entropy_14}(b) which corresponds to subtracting the sum of the entropies obtained from evolving the fermion and the antifermion wave packet individually, we see that this quantity is close to zero throughout the entire evolution. This demonstrates that there is essentially no excess entropy produced compared to evolving each wave packet individually, as expected for two noninteracting wave packets.

\subsection{Nonvanishing coupling: the interacting case}

\begin{figure*}
    \centering
    \includegraphics[width=0.8\textwidth]{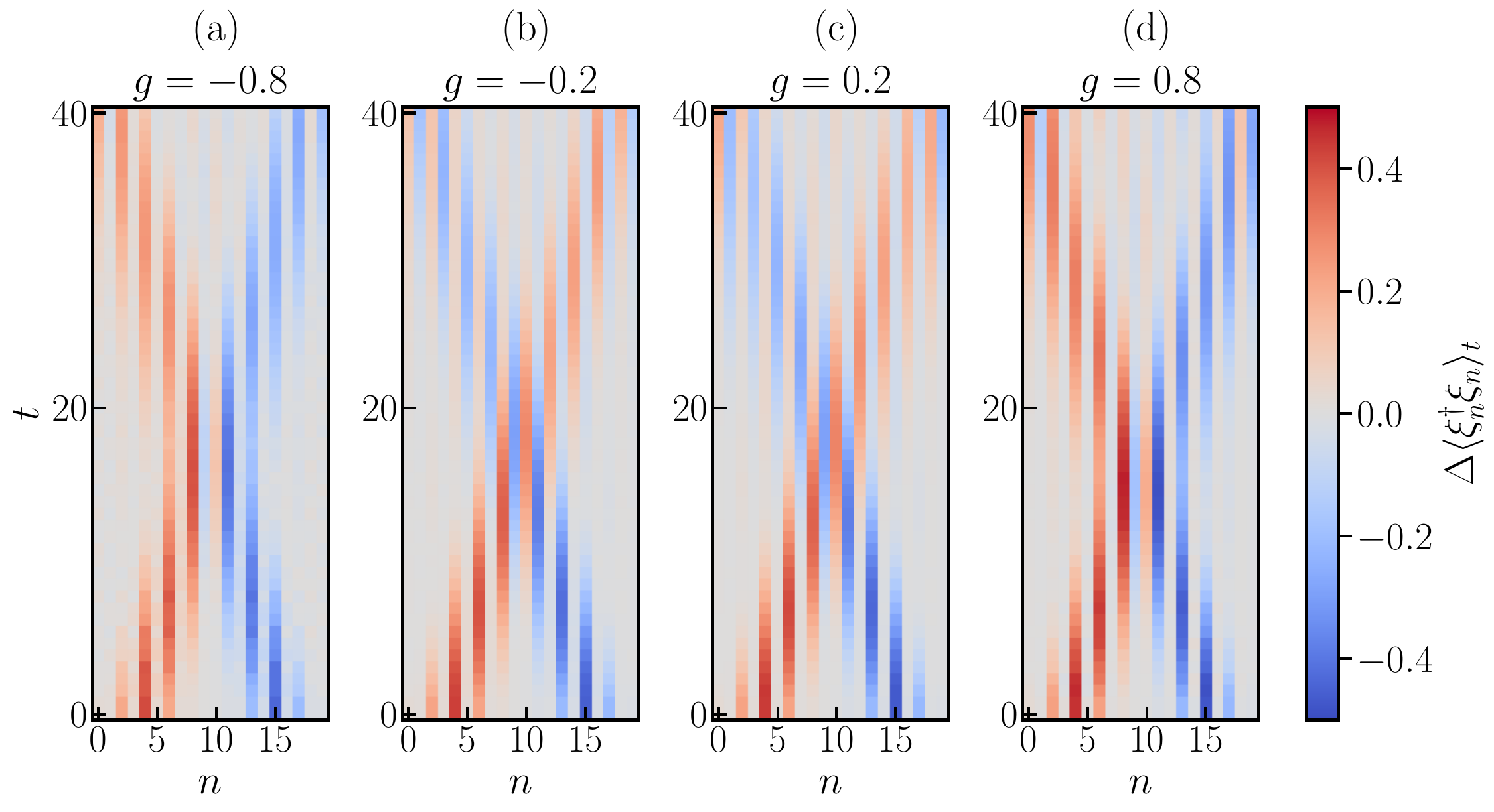}
    \caption{Site resolved particle density as a function of time for $m=1.0$ and $N=20$. The different panels correspond to different couplings (a) $g = -0.8$,  (b) $-0.2$, (c) $0.2$, and (d) $0.8$. For all simulations, the initial Gaussian fermion wave packet is located around $\mu_n^c = 4$ with mean momentum $\mu_k^c = 2\times 2\pi/N $, and the initial Gaussian antifermion wave packet is located around $\mu_n^d = 15$ with mean momentum $\mu_k^d = -2\times 2\pi/N$. Both the fermion and antifermion wave packets have a width of $\sigma_k = 2\pi/N$ in momentum space.}
    \label{fig:charge_density_interacting_fermion}
\end{figure*}

Our approach can also be applied to the interacting case, and here we study the dynamics of the lattice Thirring model with nonvanishing coupling, $g \neq 0$. To show that the operators in Eq.~\eqref{eq: givens_fermion_op} indeed create a wave packet corresponding to a particle-antiparticle pair to a good approximation, we first perform a direct evolution using a Taylor expansion up to second order as for the noninteracting case. The results for a system with $N=20$ sites and time step $\Delta t = 1\times 10^{-3}$ for various values of the coupling are shown in Fig.~\ref{fig:charge_density_interacting_fermion}. Again, we have checked that the step size is small enough to prevent any significant numerical errors. 

For $t=0$ we clearly see two wave packets corresponding to a particle and an antiparticle, again appearing respectively as an excess and a deficiency in particle density compared to the ground state; c.f. Fig.~\ref{fig:charge_density_interacting_fermion}(a). Moreover, as time progresses, we observe the wave packets are moving towards each other before eventually colliding and interacting. For small absolute values of the coupling $g$, we observe a similar behavior as in the free case: the wave packets essentially pass through each other, as shown in Figs.~\ref{fig:charge_density_interacting_fermion}(b) and \ref{fig:charge_density_interacting_fermion}(c). Comparing this to the noninteracting case (shown in Fig.~\ref{fig:charge_density_free_fermion}(c)), the wave packets disperse slightly stronger for a nonvanishing coupling of $|g|=0.2$, and the particle density spreading at $t\approx 30$ is more pronounced. At larger absolute values of the coupling, $|g|=0.8$, we observe a noticeable change in behavior. The wave packets are no longer passing through each other, but instead repel each other, as can be seen in Figs.~\ref{fig:charge_density_interacting_fermion}(a) and \ref{fig:charge_density_interacting_fermion}(d).

Our observations from the direct simulation are in agreement with the theoretical prediction for the Thirring model, with no observed particle production during the scattering process~\cite{Zamolodchikov:1977, Karowski:1977, Luscher_M:1976pb}. Due to infinite conservation laws in the Thirring model, the wave packets representing the (anti)particles either pass through each other or repel each other, depending on the coupling, $g$. While the former behavior is observed for couplings close to zero, the latter happens if the absolute value of $g$ is close to 1.

\begin{figure*}
    \centering
    \includegraphics[width=0.8\textwidth]{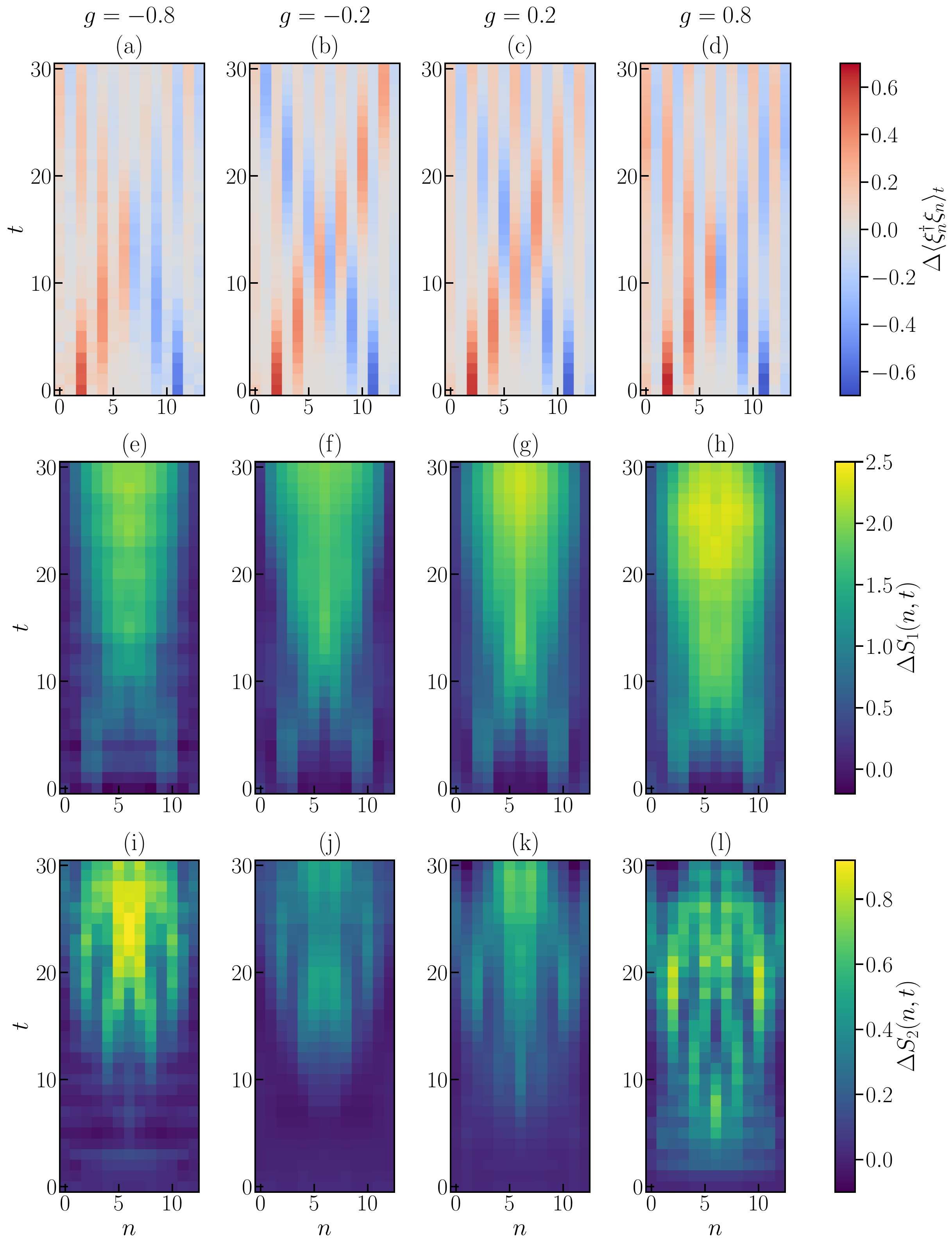}
    \caption{Particle density (first row), entropy $\Delta S_1(n,t)$ (second row) and subtracted entropy $\Delta S_2(n,t)$ (third row) for an interacting system with $14$ sites and $m=0.8$. The different columns correspond to different values of the coupling $g = \pm 0.8, \ \pm 0.2$. The initial Gaussian fermion (antifermion) wave packet is located around the site $n=2$ ($n=11$) with the mean momentum $\mu_k^c = 2\pi/ N$  ($\mu_k^d = -2\pi/ N$). Both wave packets have the same width in momentum space $\sigma_k = 2\pi/N$.}
    \label{fig:entropy_14}
\end{figure*}

Again, we also examine the von Neumann entropy a slightly smaller system size $N=14$, where we use state vector simulation in Qiskit~\cite{Qiskit} to get the vectors corresponding to the four unitary summands of Eq.~\eqref{eq:psi_evolved} and get the quantum state $|\psi(t)\rangle$ through their combination, and we normalize the state $|\psi(t)\rangle$ before calculating the von Neumann entropy. For the time evolution operator, we use the matrix representation of $e^{-iHt}$ in our simulation and the resource estimation of Trotter decomposition is left for further work. The results for various values of the coupling are depicted in Fig.~\ref{fig:entropy_14}.
Despite the reduced system size, the particle density in panels (a) - (d) clearly shows the same effect as for $N=20$: for values of the coupling close to zero the wave packets simply pass through each other, whereas for $|g|=0.8$ they repel each other. Focusing on the entropy in Fig.~\ref{fig:entropy_14}(e) - \ref{fig:entropy_14}(h), we see that there is a noticeable amount of excess entropy compared to the ground state. In particular, the entropy is growing with $|g|$, where for positive values of the coupling, we generally observe larger values of $\Delta S_1(n,t)$ than for the corresponding negative value as a comparison between Figs.~\ref{fig:entropy_14}(e) and \ref{fig:entropy_14}(h) as well as \ref{fig:entropy_14}(f) and \ref{fig:entropy_14}(g) reveals. While the values for the excess entropy compared to the noninteracting case are larger, the evolution of $\Delta S_1(n,t)$ looks qualitatively similar to the noninteracting case, as a comparison with Fig.~\ref{fig:free_entropy_14}(h) shows.

Turning towards $\Delta S_2(n,t)$ in Figs.~\ref{fig:entropy_14}(i) - \ref{fig:entropy_14}(l), we now observe a clear difference between the non-interacting case and $g\neq 0$. While the non-interacting case essentially shows a homogeneous value of  $\Delta S_2(n,t)$ throughout the entire evolution, for $g\neq 0$ we see a noticeable increase of $\Delta S_2(n,t)$ at the point where the two wave packets collide with each other. Considering the same absolute value for the coupling, we generally observe that $\Delta S_2(n,t)$ grows faster for the positive choice of $g$ than for the negative value. This is in agreement with the observation in the particle density, that for a fixed value of $|g|$, the collision happens at earlier times for a positive value than for the corresponding negative value (compare, e.g., Figs.~\ref{fig:entropy_14}(a) and \ref{fig:entropy_14}(d)). Interestingly, for the time scales we study, the data for $\Delta S_2(n,t)$ for $g=0.8$ shows a qualitatively different behavior than for $g=-0.8$. For the positive value, we see a peak in $\Delta S_2(n,t)$ in the center of the system around $t=8$ when the wave packets first start to collide. This peak subsequently decays before two peaks close to the boundary of the system emerge around $t=20$, at which time the wave packets start to interact again due to the periodic boundary conditions. In contrast, for  $g=-0.8$ we observe a considerable increase in 
$\Delta S_2(n,t)$ between the wave packets after colliding, and there is no substantial decay during the entire time we simulate. 

In summary, our numerical results demonstrate that the operators we derived allow for creating wave packets corresponding to a fermion and an antifermion with opposite momenta even in the interacting regime. The (anti)particle wave packets move towards each other and start interacting, which is reflected in $\Delta S_2(n,t)$.

\subsection{The free fermionic case on quantum hardware}
Finally, we demonstrate that our approach is feasible on current quantum hardware, and implement the noninteracting case on \texttt{ibm\_peekskill}. To this end, we use a simple parametric circuit for preparing the ground state of the model with high fidelity, where the parameters have been optimized using a classical simulation\footnote{Note that finding the optimal parameters could be done with a standard VQE approach. Since our primary focus in this work lies on the time evolution, we use a classical simulation to save time on the hardware.}. Subsequently, we evolve the initial state with two wave packets using the techniques for the free fermionic case, $g=0$, discussed in Sec.~\ref{sec_dyn} and detailed in Appendix~\ref{app: decompose_Ut_Givens}. This allows us to simulate the evolution for arbitrary times $t$ at a constant circuit depth.

In all experiments, we apply a suite of error suppression and mitigation techniques to lessen the effects of hardware noise; see Appendix~\mbox{\ref{app: qem}} for further details. To suppress decoherence and low-frequency noise on idle qubits, we apply a standard $X_p-X_m$ dynamical decoupling sequence~\mbox{\cite{PhysRev.94.630, 10.1063/1.1716296}}. To mitigate the effect of measurement errors, we apply twirled readout error extinction~\mbox{\cite{PhysRevA.105.032620}} to all observables. We employed probabilistic error amplification (PEA)~\mbox{\cite{Kim2023}} to systematically amplify the noise in our experiments and use the noise-enhanced measurements to extrapolate to the zero-noise limit, obtaining our error-mitigated observables~\mbox{\cite{PhysRevLett.119.180509, Li2017}}. For PEA, we Pauli-twirled each unique layer of two-qubit gates in our nominal circuit to transform the noise into Pauli channels. We then generated 300 circuit instances at gain factors $G = \{1, 2, 2.5\}$ and sampled each of the resulting 900 circuits 1024 times to determine the average value of each observable of interest. Finally, we used an exponential fit of the averages of the measured values at different $G$ to obtain an estimate of their zero-noise value.

\begin{figure}[htp!]
    \centering
    \includegraphics[width=0.95\columnwidth]{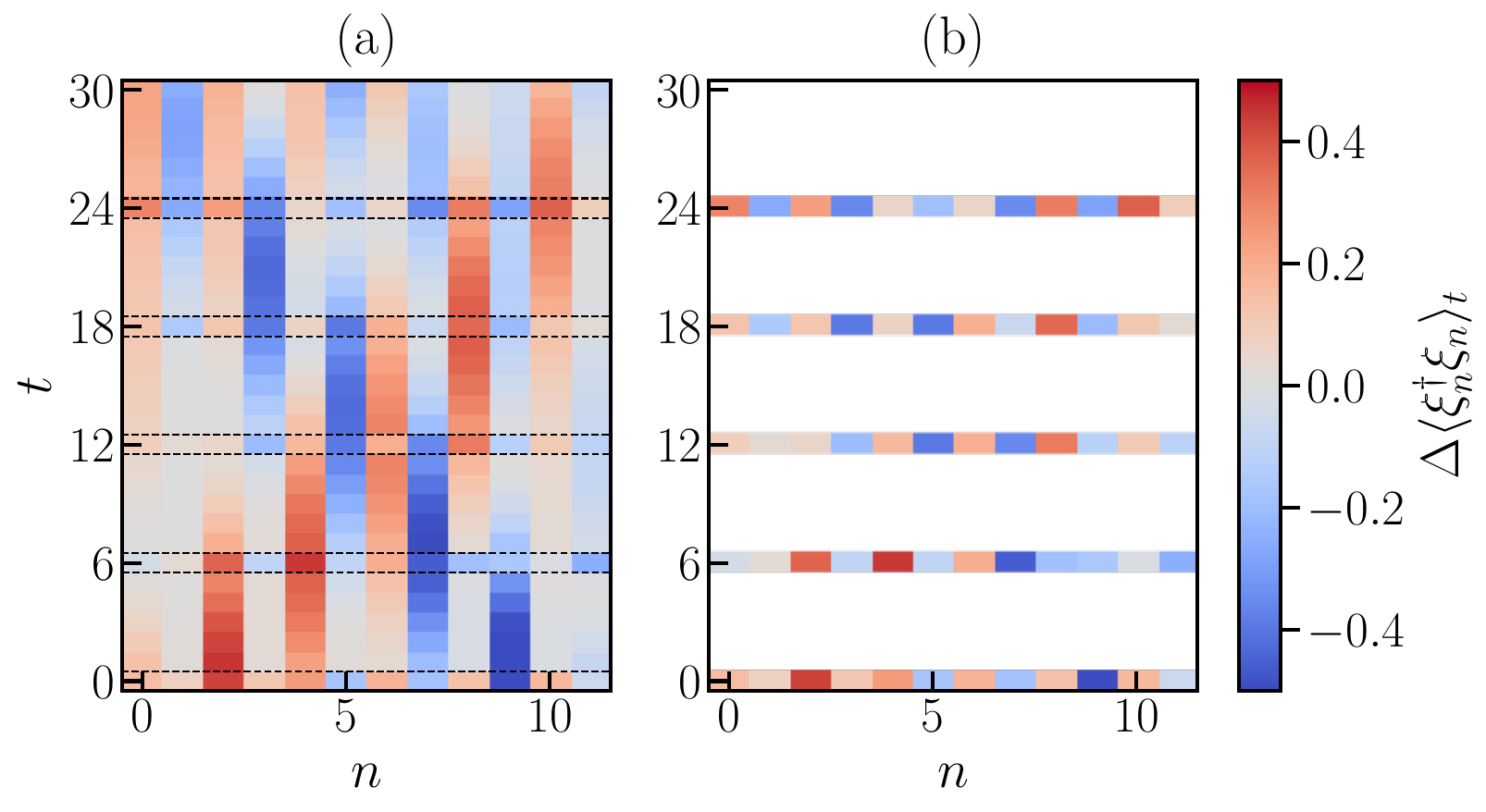}
    \caption{(a) Particle density for $N=12$ and $m=1.0$ as a function of time obtained from an ideal simulation, with the time steps at $t=0$, 6, 12, 18, 24 replaced by the data obtained from the quantum hardware (indicated by the black dashed boxes). (b) Only the data for the time slices obtained from the hardware run.}
    \label{fig:hardware_results}
\end{figure}

Figure~\ref{fig:hardware_results} shows results for the particle density obtained on \texttt{ibm\_peekskill} for 12 qubits in comparison with the exact results for several time steps. The circuit depth is up to 44 CNOT layers for all time steps, achieved by implementing the time evolution for the noninteracting case using the circuit based on Givens rotations and excluding the operators from the circuit that lie outside the light cone of the local observables, as detailed in Appendix~\ref{app: decompose_Ut_Givens}. Moreover, this allows us to directly access arbitrary time scales without having to simulate previous times. Since the computational time on the quantum hardware available to us is limited, we exploit this feature and we compute several time slices for different times instead of the entire time series.
Looking at panels \ref{fig:hardware_results}(a), we observe good agreement between the data from the quantum hardware and the exact solution. In particular, focusing on the time slices obtained from the quantum device in Fig.~\ref{fig:hardware_results}(b), we clearly see the two wave packets moving towards each other ($t=0$, 6), before they eventually pass through ($t=12$) and start to move away from each other again ($t=18$, 24). 

\begin{figure}[htp!]
    \centering
    \includegraphics[width = 0.85\columnwidth]{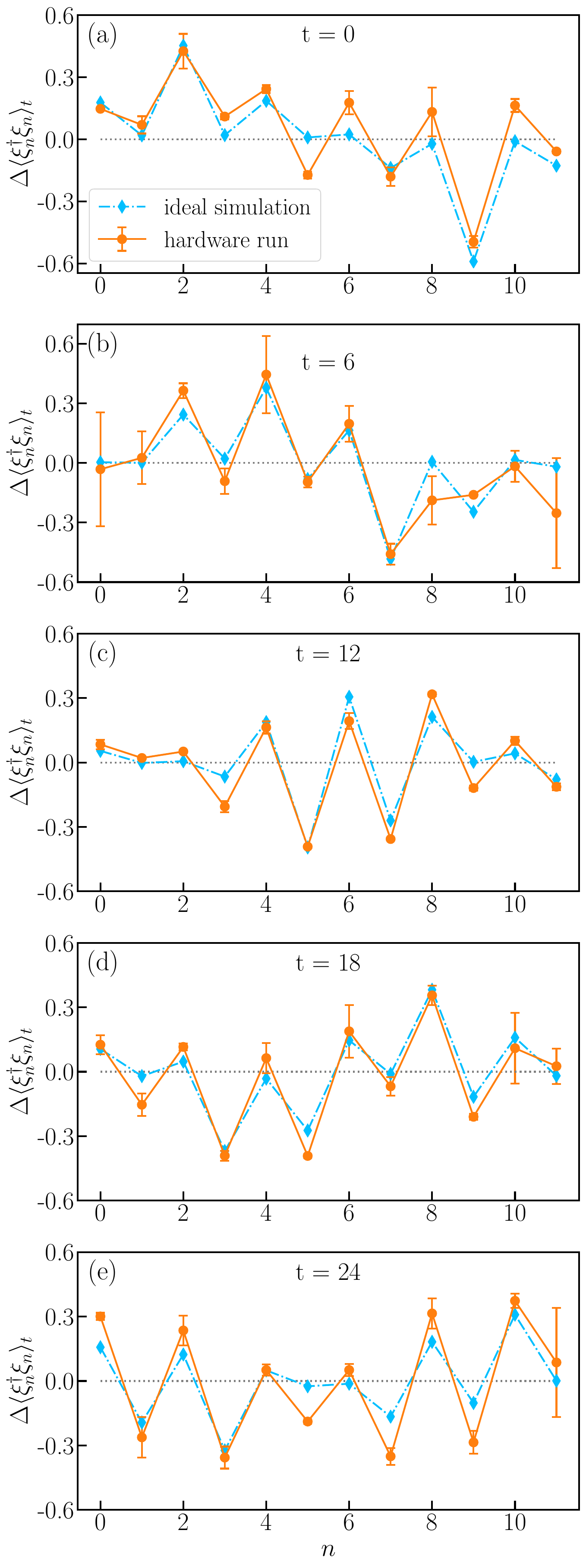}
    \caption{Time slices showing the site-resolved particle density for $N=12$ and $m=1.0$, the different panels correspond to (a) $t=0$, (b) 6, (c) 12, (d) 18, and (e) 24. The blue diamonds correspond to the ideal result on a noise-free quantum computer taking an infinite number of measurements. The yellow dots represent the data from the quantum hardware, where error bars represent uncertainties due to a finite number of measurements. As a guide for the eye, the data points are connected with lines. The horizontal dashed grey line indicates the zero value of $\Delta \langle\xi_n^\dagger\xi_n\rangle_t$.}
    \label{fig:hardware_results_individual}
\end{figure}

To get a more detailed picture of the hardware results, we show the site-resolved data for individual time slices in Fig.~\ref{fig:hardware_results_individual}.
As the figure shows, there is generally good agreement between the exact result and the experimental data from the quantum device, in most cases the analytical calculations lying within one standard deviation of the experimental uncertainty (represented by the error bars). In particular, the time-slices for the particle density illustrate once more the presence of two separated wave packets, showing up as a peak in the particle density around site $2$ and a dip at site $9$ for $t=0$ (c.f.\ Fig.~\ref{fig:hardware_results_individual}(a)). These are moving towards the center as time progresses to $t=12$,  resulting in a particle density noticeably different from zero in the center of the system, as shown in Fig.~\ref{fig:hardware_results_individual}(c). Eventually, after the two wave packets pass through each other, we observe again a well separated peak and a dip in the particle density around $t=18$ in Fig.~\ref{fig:hardware_results_individual}(d). 

\section{Conclusion and outlook\label{sec:conclusion}}

In this work, we proposed a framework for studying fermion scattering on digital quantum computing. Using the lattice Thirring model as an example, we demonstrated our framework by simulating the elastic collision for fermion-antifermion wave packets both classically and on quantum hardware.

Guided by the free theory, corresponding to the Thirring model at vanishing coupling, we derived a set of operators that allow us to create approximate fermion and antifermion wave packets for the interacting theory on top of the ground state. Starting from such an initial state, we showed how to efficiently obtain the expected value of observables from a quantum device throughout the evolution and provided the necessary quantum circuits to measure them.

To demonstrate our approach, we first simulated the dynamics of a fermion-antifermion wave packet in the free theory exactly before proceeding to the interacting case. Observing the particle density and the von Neumann entropy produced throughout the elastic scattering, we characterized the process and showed that our framework provides an avenue towards simulating these dynamics on quantum devices. While the entropy can in general not be obtained efficiently on a quantum device, the particle density can be readily measured on a digital quantum computer. Moreover, we carried out a proof-of-principle demonstration simulating the scattering of a fermion and an antifermion wave packet for the free theory on IBM's quantum devices. Using state-of-the-art error mitigation methods, the data obtained from the quantum device is in good agreement with the theoretical prediction, thus showing that our approach is suitable for near-term quantum hardware.

Our work can be generalized to arbitrary purely fermionic models that have a parameter point at which they are at most quadratic in the fermionic operators, and provides a first step towards simulating fermionic scattering processes on quantum computers. While we provided a demonstration on quantum hardware, a systematic study of the effects of the quantum noise is left for future work. Furthermore, our data for the particle density can serve as a benchmark for future experiments studying scattering processes in the Thirring model on a quantum device.

\begin{acknowledgments}
    The authors thank the QC4HEP working group for useful discussions. IBM, the IBM logo, and ibm.com are trademarks of International Business Machines Corp., registered in many jurisdictions worldwide. Other product and service names might be trademarks of IBM or other companies. The current list of IBM trademarks is available at \url{https://www.ibm.com/legal/copytrade}.
    S.K.\ acknowledges financial support from the Cyprus Research and Innovation Foundation under the project ``Quantum Computing for Lattice Gauge Theories (QC4LGT)'', contract no.\ EXCELLENCE/0421/0019.
    A. C. is supported in part by the Helmholtz Association —“Innopool Project Variational Quantum Computer Simulations (VQCS).”
    This work is funded by the European Union's Horizon Europe Framework Programme (HORIZON) under the ERA Chair scheme with grant agreement no.\ 101087126.
    This work is supported with funds from the Ministry of Science, Research and Culture of the State of Brandenburg within the Centre for Quantum Technologies and Applications (CQTA). 
    \begin{center}
        \includegraphics[width = 0.08\textwidth]{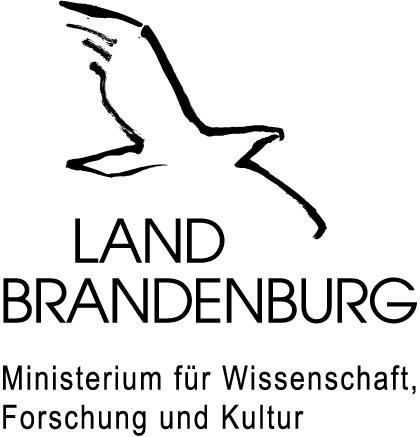}
    \end{center}
\end{acknowledgments}

Data available to anyone upon a reasonable request.

\appendix
\widetext
\section{Details of Givens rotation\label{app:Givens_rotation}}
In this section, we explain how to decompose a complex unitary matrix $u$ based on Givens rotation. We then show how to decompose the operation $V(u)$ into local operations based on this decomposition of $u$. 

Given a complex unitary matrix $u$ with dimensions $N \times N$, we can find a series unitary matrices $r_{n,0}(\theta_{n,0})$ and $p_l(\vec{\beta}_l)$ that diagonalize the matrix $u$ as follows
\begin{align}\label{eq: s}
    &s_{N-1} \cdots s_0  u = I,
\end{align}
where 
\begin{align}\label{eq: s_rp}
    &s_l = r_{l+1,l}(\theta_{l+1,l}) \cdots r_{N-1,l}(\theta_{N-1,l})  \cdot p_l(\vec{\beta}_l), \ \  l = 0, \cdots N-2, \\
    &s_{N-1} = p_{N-1}(\vec{\beta}_{N-1}).
\end{align}
The matrix $s_l$ serves to diagonalize the $l$-th column of matrix $u$. In this process, the diagonal matrix $p_l(\vec{\beta}_l)$ eliminates the phase of elements in column $l$ and then the Givens rotation matrix $r_{n,l}(\theta)$ eliminates the element $u_{n,l}$. The explicit expression for the matrices $p_l(\vec{\beta}_l)$ and $r_{n,l}(\theta)$ read
\begin{align}
    p_l(\vec{\beta}_l)=
    \begin{pmatrix}
 e^{i\beta_{0,l}} & 0 &\cdots &0\\
 0 & e^{i\beta_{1,l}} &\cdots &0\\
 \vdots& \vdots &\ddots &\vdots \\
0 & \cdots &\cdots &e^{i\beta_{N-1,l}}
\end{pmatrix}_{N \times N},\quad
    r _{n,l}(\theta)=
    \begin{pmatrix}
 &1 & \cdots & 0 & 0 &\cdots &0\\
 &\vdots & \ddots & \vdots &\vdots & & \vdots\\
 &0 & \cdots & \cos(\theta) & -\sin(\theta) & \cdots &0 \\
 &0 & \cdots & \sin(\theta) & \cos(\theta) & \cdots &0 \\
&\vdots & & \vdots & \vdots &\ddots &\vdots \\
&0 & \cdots &0 &0 & \cdots & 1
\end{pmatrix}_{N \times N},
\end{align}
where the matrix $r_{n,l}(\theta)$ is diagonal, except for the $(n-1)$-th and $n$-th rows and columns. 

Below, we illustrate the procedure for the diagonalization of the first column of $u$. Considering a general complex unitary matrix $u$,
\begin{equation}
    u =
    \begin{pmatrix}
 a_{0,0} e^{-i b_{0,0}} & a_{0,1} e^{-i b_{0,1}} &\cdots & a_{0,N-1} e^{-i b_{0,N-1}}\\
 \vdots                 &  \vdots                &\cdots & \vdots\\
 a_{N-2, 0} e^{-i b_{N-2, 0}} & a_{N-2, 1} e^{-i b_{N-2, 1}} &\cdots & a_{N-2, N-1} e^{-i b_{N-2, N-1}}\\
 a_{N-1, 0} e^{-i b_{N-1, 0}} & a_{N-1, 1} e^{-i b_{N-1, 1}} &\cdots & a_{N-1, N-1} e^{-i b_{N-1, N-1}} \\
\end{pmatrix}_{N \times N},
\end{equation}
 where $a_{n,l}$, $b_{n,l}$ represent the absolute value and the phase of the matrix element $u_{n,l}$. In order to diagonalize the first column of $u$, we will multiply the $p_0(\vec{\beta}_0)$ to $u$ such that the elements in the first column of $u$ become real. The resulting matrix $p_0(\vec{\beta}_0) \cdot u$ is given by
 \begin{equation}
    p_0(\vec{\beta}_0) \cdot u =
    \begin{pmatrix}
 a_{0,0} & a_{0,1} e^{-i \widetilde{b}_{0,1}} &\cdots & a_{0,N-1} e^{-i \widetilde{b}_{0,N-1}}\\
 \vdots                 &  \vdots                &\cdots & \vdots\\
 a_{N-2, 0} & a_{N-2, 1} e^{-i \widetilde{b}_{N-2, 1}} &\cdots & a_{N-2, N-1} e^{-i \widetilde{b}_{N-2, N-1}}\\
 a_{N-1, 0} & a_{N-1, 1} e^{-i \widetilde{b}_{N-1, 1}} &\cdots & a_{N-1, N-1} e^{-i \widetilde{b}_{N-1, N-1}} \\
\end{pmatrix}_{N \times N}.
\end{equation}
In the expression above, the phase of the first column in $u$ is eliminated by $p_0(\vec{\beta}_0)$ with $\beta_{n,0}=b_{n,0}$, and the phase of elements in other columns are changed to $\widetilde{b}_{n,l} = b_{n,l} - \beta_{n,0}$. Subsequently, we can utilize  the matrix $r_{N-1,0}(\theta_{N-1,0})$ to induce a rotation between the $(N-1)$-th row and $(N-2)$-th row. Choosing the angle $\theta_{N-1,0} = \arctan{ \left(-a_{N-1,0}/a_{N-2,0}\right)}$, the rotation is able to transform the element $a_{N-1, 0}$ into a zero
\begin{equation}
    \begin{aligned}
        &r_{N-1,0}(\theta_{N-1,0}) \cdot p_0(\vec{\beta}_0) \cdot u  \\
        &=
        \begin{pmatrix}
         a_{0,0} & a_{0,1} e^{-i \widetilde{b}_{0,1}} &\cdots & a_{0,N-1} e^{-i \widetilde{b}_{0,N-1}}\\
         \vdots                 &  \vdots                &\ddots & \vdots\\
         \cos(\theta_{N-1,0})a_{N-2, 0} - \sin(\theta_{N-1,0}) a_{N-1,0} & \widetilde{a}_{N-2, 1} e^{-i \widetilde{b}_{N-2, 1}} &\cdots & \widetilde{a}_{N-2, N-1} e^{-i \widetilde{b}_{N-2, N-1}}\\
          \cos(\theta_{N-1,0})a_{N-1, 0} + \sin(\theta_{N-1,0}) a_{N-2,0} & \widetilde{a}_{N-1, 1} e^{-i \widetilde{b}_{N-1, 1}} &\cdots & \widetilde{a}_{N-1, N-1} e^{-i \widetilde{b}_{N-1, N-1}} \\
        \end{pmatrix}_{N \times N} \\
        &\stackrel{\theta_{N-1,0} = \arctan{ \left(-\frac{a_{N-1,0}}{a_{N-2,0}}\right)}}{\longrightarrow}
            \begin{pmatrix}
         a_{0,0} & a_{0,1} e^{-i \widetilde{b}_{0,1}} &\cdots & a_{0,N-1} e^{-i \widetilde{b}_{0,N-1}}\\
         \vdots                 &  \vdots                &\ddots & \vdots\\
         \widetilde{a}_{N-2, 0} & \widetilde{a}_{N-2, 1} e^{-i \widetilde{b}_{N-2, 1}} &\cdots & \widetilde{a}_{N-2, N-1} e^{-i \widetilde{b}_{N-2, N-1}}\\
          0 & \widetilde{a}_{N-1, 1} e^{-i \widetilde{b}_{N-1, 1}} &\cdots & \widetilde{a}_{N-1, N-1} e^{-i \widetilde{b}_{N-1, N-1}} \\
        \end{pmatrix}_{N \times N}
    \end{aligned}
    \label{eq:first_steps_givens_roation}
\end{equation}
In the matrix in Eq.~\eqref{eq:first_steps_givens_roation}, all the matrix elements in the $(N-1)$-th row and the $(N-2)$-th row change due to the Givens rotation. To simplify the notation, we use $\widetilde{a}_{n,l}$ and $\widetilde{b}_{n,l}$ to represent the absolute value and phase of matrix elements after the rotation. To fully diagonalize the first column, we need to multiply more Givens rotation matrices to $u$ in order to set the remaining entries in the rows $N-2, \dots , 1$ to zero
\begin{equation}
 \begin{aligned}
    & s_0 \cdot u = r_{1,0}(\theta_{1,0}) \cdot \dots \cdot r_{N-1,0}(\theta_{N-1,0}) \cdot p_0(\vec{\beta}) \cdot u  \\
&\stackrel{\theta_{n,0} =\arctan{ \left(-\frac{a_{n,0}}{a_{n-1,0}}\right) }}{\longrightarrow}
    \begin{pmatrix}
 1     &    0     &\cdots & 0\\
 0     &    \widetilde{a}_{1,1} e^{-i \widetilde{b}_{1,1}}    &\cdots & \widetilde{a}_{1,N-1} e^{-i \widetilde{b}_{1,N-1}}\\
 \vdots&  \vdots  &\ddots & \vdots\\
 0 & \widetilde{a}_{N-2, 1} e^{-i \widetilde{b}_{N-2, 1}} &\cdots & \widetilde{a}_{N-2, N-1} e^{-i \widetilde{b}_{N-2, N-1}}\\
  0 & \widetilde{a}_{N-1, 1} e^{-i \widetilde{b}_{N-1, 1}} &\cdots & \widetilde{a}_{N-1, N-1} e^{-i \widetilde{b}_{N-1, N-1}} \\
\end{pmatrix}_{N \times N},
\end{aligned}
\end{equation}
where the angles $\theta_{n,0} $ are again given by $\arctan{ \left(-a_{n,0}/a_{n-1,0}\right)}$ for $n = N-1, N-2, \cdots ,1$. Furthermore, the first elements of other columns are also zero after diagonalizing the first column. This is a consequence of the orthogonality between different columns in a unitary matrix. To fully diagonalize $u$, one can carry out the same procedure for the remaining columns. 

In our wave packet preparation procedure, we specifically focus on the first column of $u$, denoted by $u_{n,0} = \tilde{\phi}_n^{c(d)}$. Therefore, assuming $s_0 \cdot u = I$, the decomposition of $u$ is given by
\begin{equation}
   u = (s_0)^{\dagger} = \left( p_0(\vec{\beta}_0) \right)^{\dagger} \cdot \left(r_{N-1,0}(\theta_{N-1,0}) \right)^{\dagger} \cdot \dots \cdot \left(r_{1,0}(\theta_{1,0})\right)^{\dagger}.
\end{equation}
Regarding the unitary operator $V(u)$ defined in Eq.~\eqref{eq: V_u}, one can get a decomposition of it based on the property $V(u \cdot u^{\prime}) = V(u) \cdot V(u^{\prime})$ and the decomposition of matrix $u$:
\begin{equation}
    V(u) = V^{\dagger}(p_0) \cdot V^{\dagger}(r_{N-1,0}) \cdots V^{\dagger}(r_{1,0}), \\
\end{equation}
with,
\begin{equation}
\begin{aligned}
   & V^{\dagger}(p_0) = \exp\left( -\sum_{nl} [\log{p_0}]_{nl}\  \xi_n^{\dagger} \xi_l\right) = \exp\left( -i\sum_n \beta_{n,0}\  \xi_n^{\dagger} \xi_n\right), \\
   & {V}^{\dagger}(r_{n,0}) = \exp\left( -\sum_{ij} [\log{r_{n,0}}]_{ij}\  \xi_i^{\dagger} \xi_j\right) = \exp\left( \theta_{n,0}  [\xi^{\dagger}_{n-1} \xi_n - \xi^{\dagger}_n \xi_{n-1}] \right).
\end{aligned}
\end{equation}
The final equation above uses that $\log { r_{n,l}(\theta) }$ has non-zero entries only for the elements with indices $(i,j) = (n-1, n)$ and $(n, n-1)$
\begin{equation}
   \log { r_{n,l}(\theta) }=
    \begin{pmatrix}
        0      & \cdots & 0      & 0       & \cdots & 0 \\
        \vdots & \ddots & \vdots & \vdots  &        & \vdots \\
        0      & \cdots & 0      & -\theta & \cdots & 0 \\
        0      & \cdots & \theta & 0       & \cdots & 0 \\
        \vdots &        & \vdots & \vdots  &\ddots  & \vdots \\
        0      & \cdots & 0      & 0       & \cdots & 0
    \end{pmatrix}_{N \times N}.
\end{equation}

\section{Momentum space and Fock space\label{app:momentum_and_fock_space}}
In this appendix, we analytically determine the particle density for the noninteracting case during the scattering process via calculations in momentum and Fock space. Since we are dealing with a free fermionic system, these computations involve matrices whose dimensions $N \times N$, enabling us to numerically simulate the phenomenon for large system and to cross-check the results in position space.

In momentum space, the free fermion Hamiltonian is diagonal in terms of the fermionic and antifermionic creation (annihilation) operators $c_k^{\dagger}$, $d_k^{\dagger}$ ($c_k$, $d_k$)~\cite{Simone_prd}
\begin{align}\label{eq:H_momentum}
    H = \sum_k w_k \left( c_k^{\dagger}c_k - d_k d_k^{\dagger} \right),\quad k \in \Lambda_k = \frac{2\pi}{L} \left\{-\left\lfloor \frac{N}{4} \right\rfloor, \cdots \left\lceil \frac{N}{4} \right\rceil-1  \right\},
\end{align}
and the vacuum $\ket{\upOmega}$ in this theory is the state with no excitations of fermions and antifermions:
\begin{equation}\label{eq: annihilation_vaccum}
    c_k \ket{\upOmega} = d_k \ket{\upOmega} = 0.
\end{equation}
As there is no interaction between fermions and antifermions in momentum space, the wave function of the system is defined as the tensor product of the wave function for fermion and antifermion subsystems:
\begin{equation}
    \ket{\psi} = \ket{\psi_c} \otimes \ket{\psi_d}.
\end{equation}

\begin{figure}[htp!]
    \centering
    \includegraphics[width = 0.4\textwidth]{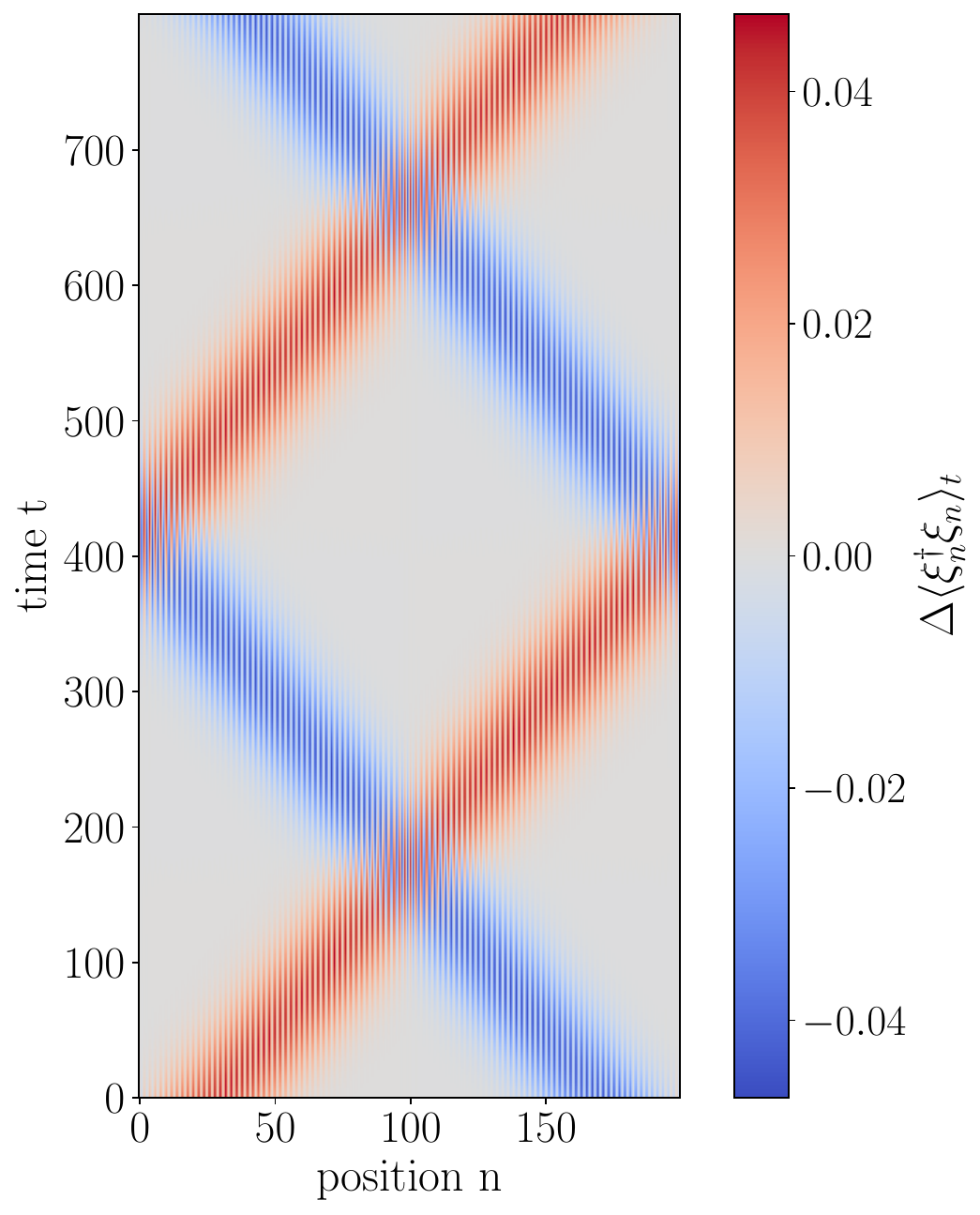}
    \caption{The particle density for free staggered fermions on a lattice with $N = 200$ sites. The initial fermionic Gaussian wave packet is located at $\mu_n^c = 30$ with mean momentum $\mu_k^c = 20 \times 2\pi/N$; the initial antifermion wave packet is located at $N-1-\mu_n^c$ with the same absolute value for the momentum as the fermion but in the opposite direction. For both the fermion and antifermion wave packet we choose a width of $\sigma_k = 2\pi/N$ in momentum space.}
    \label{fig:N_200}
\end{figure}

When considering the Fock basis for the momentum state, the Gaussian wave packets for the fermion and antifermion can be represented as the vectors 
\begin{align}
    \ket{\psi_c(t=0)} &\rightarrow \left(\phi^c_{k_1}, \dots , \phi^c_{k_{N/2}}  \right)^\mathrm{T}, \\
    \ket{\psi_d(t=0)} &\rightarrow \left(\phi^d_{k_1}, \dots , \phi^d_{k_{N/2}}  \right)^\mathrm{T},
\end{align}
with $k_i = 2\pi/N \times \left( i - N/4 - 1 \right)$. In this basis, the fermion and antifermion part of the Hamiltonian correspond to diagonal matrices with dimension $N/2$ and the full Hamiltonian will be a direct sum of the matrix for the fermions and the antifermions. The fermionic part will be given by
\begin{align}
    H_c &= \sum_k w_k c_k^{\dagger} c_k
    \rightarrow 
    \begin{pmatrix}
        w_0    & 0      & \cdots & 0 \\
        0      & w_2    & \cdots & 0 \\
        \vdots & \vdots & \ddots & \vdots\\
         0     & 0      & \cdots & w_{N/2-1}
    \end{pmatrix}
\end{align}
and analogously the antifermionic part. As a result, the time evolution of the initial fermionic Gaussian wave packet can be expressed as follows (up to a global phase)
\begin{align}\label{eq: time_evolution_momentum}
    \ket{\psi(t)} &= e^{-iHt} \ket{\psi(0)}= \left( e^{-iH_c t} \ket{\psi_c(0)} \right) \otimes \left( e^{-iH_d t} \ket{\psi_d(0)} \right) \\
    &\rightarrow \left(e^{-iw_{k_1}t}\phi^c_{k_1}, \dots , e^{-iw_{k_{N/2}}t}\phi^c_{k_{N/2}}  \right)^\mathrm{T} \otimes 
    \left(e^{-iw_{k_1}t}\phi^d_{k_1}, \dots , e^{-iw_{k_{N/2}}t}\phi^d_{k_{N/2}}  \right)^\mathrm{T},
\end{align}
To obtain the particle density $\langle \xi_n^{\dagger} \xi_n \rangle_t$ at each spatial site, we need to transform the operator $\xi_n^{\dagger} \xi_n$ into momentum space. For even $n$ we find
\begin{align}\label{eq: xidxi_even}    
    \begin{aligned}
        \xi_n^{\dagger} \xi_n &= \frac{1}{N} \sum_{pq\in \Lambda_k} \sqrt{\frac{m+w_p}{w_p}}\sqrt{\frac{m+w_q}{w_q}} e^{-i(p-q) n} \times c_p^{\dagger} c_q\\
        &- \frac{1}{N} \sum_{pq\in \Lambda_k} \sqrt{\frac{m+w_p}{w_p}}\sqrt{\frac{m+w_q}{w_q}} e^{-i(p-q) n} \times d_p^{\dagger} d_q \times  v_p v_q + \frac{1}{N} \sum_{p} {\frac{m+w_p}{w_p}} \times v_p^2, \\
        &\rightarrow \sum_{pq\in \Lambda_k} \left( c_p^{\dagger} \cdot C^{n0}_{pq} \cdot c_q + d_p^{\dagger} \cdot D^{n0}_{pq} \cdot d_q \right) + C_\text{even}, 
    \end{aligned}
\end{align}   
For odd $n$, we obtain
\begin{align}\label{eq: xidxi_odd}
    \begin{aligned}
        \xi_n^{\dagger} \xi_n &= \frac{1}{N} \sum_{pq\in \Lambda_k} \sqrt{\frac{m+w_p}{w_p}}\sqrt{\frac{m+w_q}{w_q}} e^{-i(p-q) n} \times c_p^{\dagger} c_q \cdot  v_p v_q\\
        &- \frac{1}{N} \sum_{pq\in \Lambda_k} \sqrt{\frac{m+w_p}{w_p}}\sqrt{\frac{m+w_q}{w_q}} e^{-i(p-q) n} \times d_p^{\dagger} d_q + \frac{1}{N} \sum_{p} {\frac{m+w_p}{w_p}}, \\
        &\rightarrow \sum_{pq\in \Lambda_k} \left( c_p^{\dagger} \cdot C^{n1}_{pq} \cdot c_q + d_p^{\dagger} \cdot D^{n1}_{pq} \cdot d_q \right) + C_\text{odd},
    \end{aligned}
\end{align}
where the constants $C_\text{even} = \frac{1}{N} \sum_{p} {\frac{m+w_p}{w_p}} \times v_p^2, \ C_\text{odd} = \frac{1}{N} \sum_{p} {\frac{m+w_p}{w_p}}$, which are the particle density of the vacuum for even and odd sites, e.g., $\bra{\upOmega} \xi_n^{\dagger} \xi_n \ket{\upOmega} = C_\text{even}\ \text{or}\  C_\text{odd}$ according to Eq.~\eqref{eq: annihilation_vaccum}. Note that $C_\text{even} + C_\text{odd} = 1$, which means the vacuum of free fermion has particle number $\sum_n \bra{\upOmega} \xi_n^{\dagger} \xi_n \ket{\upOmega} = \left( C_\text{even} + C_\text{odd} \right) \times N/2= N/2$. The definitions of $w_k$ and $v_k$ can be found in Eq.~\eqref{eq:wk_vk}. $C^{nr}_{pq}, D^{nr}_{pq}, r\in \{0, 1\}$ are coefficients of the operator $c_p^{\dagger} c_q$ or $d_p^{\dagger} d_q$ in the first line of Eq.~\eqref{eq: xidxi_even} and Eq.~\eqref{eq: xidxi_odd}. Using the matrix representation in one-particle Fock space $C^{nr}, D^{nr}, r\in \{0, 1\}$, we can evaluate the particle density that deviate from vacuum quickly by  calculating $\vec{\phi_k^c}^T \cdot C^{nr} \cdot \vec{\phi_k^c}  + \vec{\phi_k^d}^T \cdot D^{nr} \cdot \vec{\phi_k^d} $. We show the result for 200 sites in Fig.~\ref{fig:N_200}.

\section{Details on the classical numerical simulations\label{app:classical_numerics}}

This appendix provides some details regarding the classical numerical simulations used to the generate the results in Sec.~\ref{sec:results}.

For the results for $N=20$, shown in Figs.~\ref{fig:charge_density_free_fermion} and \ref{fig:charge_density_interacting_fermion}, we first compute the ground state using standard Krylov methods. Subsequently, we apply the operators $C^\dagger$ and $D^\dagger$ to the ground state and simulate the time evolution by approximating the the exact time evolution operator, via a second order Taylor series expansion, $\exp(-iH\Delta t) \approx 1- i\Delta t H -  H^2 (\Delta t)^2/2$. This approach is chosen because computing the exact time evolution operator generally requires fully diagonalizing the Hamiltonian, and results in a dense matrix, even if $H$ is sparse. For 20 qubits storing the dense matrix representing the time evolution operator would require around 18Tb of memory. In contrast, using the second order Taylor expansion, we only have to apply the sparse representation of $H$ to a vector at each step, which requires a lot less memory. For finite $\Delta t$, this approximation does not preserve the unitarity of the evolution operator and the wave function has to be renormalized after every step. We checked that our choice of $\Delta t = 2\times 10^{-3}$ does not introduce any noticeable errors during the time scales we simulate.

For $N=14$, we simulate the time-evolution operator $e^{-iHt}$ exactly in both the non-interacting and interacting case using Qiskit's state vector simulator backend. We proceed in two different ways, depending if we are treating the noninteracting or the interacting case. For the noninteracting case, we are simulating the circuits we derived in Secs.~\ref{sec_dyn} and \ref{sec_td_obs} and directly implement these in Qiskit. For the interacting case, we define $e^{-iHt}$ as a gate operation in Qiskit using the matrix representation. In all of these cases, the ground state has been obtained from exact diagonalization of Hamiltonian and is encoded in the circuit using Qiskit functionality to encode a vector into a circuit.  Again,  for both cases, the vacuum state has been obtained from exact diagonalization of Hamiltonian.

To compute the von Neumann entropy, we have to obtain the reduced density operator $\rho_n$. Even if $\ket{\psi(t)}$ is a pure state, $\rho_n$ is in general a dense matrix of dimension $2^n\times 2^n$. The von Neumann entropy can then be computed from the spectrum $\eta_i$, $i=1,\dots,2^n$ of $\rho_n$, which can be determined via exact diagonalization. Given the eigenvalues of the reduced density operator, the von Neumann entropy can be calculated via by $S=-\sum_{i=1}^{2^n} \eta_i \log_2 \eta_i$. In our simulations, we consider all possible  bipartions $n=1,\dots,N-1$. Hence, we reduce system sizes to $N=14$ in case we want to compute the von Neumann entropy to avoid similar memory issues as described for the exact computation of the time-evolution operator.

\section{Hardware run for the noninteracting case\label{app: decompose_Ut_Givens}}

As shown in Sec.~\ref{sec_circuits}, the initial state of the scattering process that consists of fermionic wave packets can be prepared by the quantum circuit corresponding to Givens rotation. For the special case of having exactly one fermion wave packet and one antifermion wave packet, the ansatz can be simplified by utilizing the orthogonality of the vectors $\phi^c$ and $\phi^{d\ast}$, $\sum_n \tilde{\phi}_n^d  \tilde{\phi}_n^c = 0$ (see Appendix~\ref{app: orthogonality}). In this case the amplitudes $\phi^c_n$ and $\phi^{d\ast}$ can be regarded as the first and second column of a unitary matrix $u_{cd}$, and the operators $C^{\dagger}(\tilde{\phi}_n^c), D^{\dagger}(\tilde{\phi}_n^d)$ can be expressed as
\begin{equation}\label{eq:CD_operators}
    \begin{aligned}
        C^{\dagger}(\tilde{\phi}_n^c) &= V(u_{cd})\xi_0^\dagger V^\dagger(u_{cd}) = \sum_n \xi^\dagger_n u_{n,0} = \sum_n \tilde{\phi}_n^c \xi_n^{\dagger}, \\
        D^{\dagger}(\tilde{\phi}_n^d) &= V(u_{cd})\xi_1 V^\dagger(u_{cd}) = \sum_n \xi_n u_{n,1}^{\ast} = \sum_n \tilde{\phi}_n^d \xi_n. \\
    \end{aligned}
\end{equation}
Using Eq.~\eqref{eq:CD_operators}, the initial state of the scattering process can be prepared as
\begin{equation}
    \begin{aligned}
        \ket{\psi(0)} &= D^{\dagger}(\tilde{\phi}_n^d) \cdot C^{\dagger}(\tilde{\phi}_n^c)\ket{\upOmega} \\
        &= \left( V(u_{cd}) \xi_1 V^\dagger(u_{cd}) \right) \cdot \left( V(u_{cd}) \xi_0^\dagger V^\dagger(u_{cd}) \right) \ket{\upOmega} \\
        &= V(u_{cd})\sigma_1^+ \sigma_0^z \cdot \sigma_0^- V^\dagger(u_{cd})\ket{\upOmega},
    \end{aligned}
\end{equation}
where in the step from the second to the last line we have mapped the fermion operators to Pauli operators using a Jordan-Wigner transformation, and used that $V(u)$ is unitary. This approach gives the same result as in Eq.~\eqref{eq: givens_fermion_op}, but with a shallower circuit. Note that the method described above for simplifying the circuit for wave packet preparation is not applicable for states consisting of either two fermions or two antifermions, due to the non-orthogonality of their respective coefficients, a detail explained in Appendix~\ref{app: orthogonality}. Additionally, in the noninteracting case, the nonunitary operator $\sigma_1^+, \sigma_0^-$ in the last line can be substituted by $\sigma_1^x, \sigma_0^x$, as in Eq.~\eqref{eq:psi_init_free_case}.

Furthermore, in the noninteracting case, corresponding to $g = 0$ in Eq.~\eqref{eq:hamiltonian_lattice_fermionic}, the Hamiltonian is quadratic in terms of fermionic operators and thus corresponds to a free staggered Dirac fermion.  The time evolution operator for this case reads as
\begin{equation}
    \begin{aligned}
        U(t) &= \exp \left( -i H t \right) \\
         &= \exp \left( -i t \sum_{n=0}^{N-1} \left( \frac{i}{2} \left( \xi_{n+1}^{\dagger}\xi_{n} - \xi_n^{\dagger}\xi_{n+1}\right) + (-1)^n m \, \xi_n^{\dagger}\xi_n \right) \right) \\
         &= \exp \left( \sum_{nl} \xi_n^{\dagger}  M_{nl} \xi_l \right),
    \end{aligned}
    \label{eq: Ut_free_fermion}
\end{equation}
where $M$ is an antihermitian matrix in Fock space depending only on the mass and evolution time
\begin{equation}
    M = t \times
    \begin{pmatrix}
    -im & -1/2 &   0  &   0  &\cdots & 0 &1/2\\
    1/2 & im   & -1/2 &   0  & \cdots& 0 &0 \\
     0  & 1/2  & -im  & -1/2 & \cdots& 0 &0 \\
  \vdots&\vdots&\vdots&\vdots&\ddots &\vdots&\vdots\\
   -1/2 &  0   &  0   &   0  &\cdots &1/2 & im
\end{pmatrix}_{N \times N}.
\end{equation}

As Eq.~\eqref{eq: Ut_free_fermion} reveals, the operator $U(t)$ has the same form as Eq.~\eqref{eq: V_u} with $u = u_t := e^M$. Thus, we can use Givens rotations to decompose the time-evolution operator in the same way as $V(u)$. Appendix~\ref{app:Givens_rotation} provides a comprehensive explanation of how to use Givens rotation to diagonalize the first column of the unitary matrix $u$. The remaining columns can then be successively brought to diagonal form by repeating the procedure, eventually resulting in the decomposition of $U(t)$. Thus, for the noninteracting case, the time-evolved state of the scattering process can be completely prepared by Givens rotation 
\begin{equation}\label{eq: psit_free}
    \begin{aligned}
    \ket{\psi(t)} &= e^{-iHt} \ket{\psi(0)} \\
    &=V(u_t) \cdot V(u_{cd})\sigma_1^x (i\sigma_0^y) V^\dagger(u_{cd}) \ket{\upOmega}\\
    &=V(u_t^{\prime}) \cdot \sigma_1^x (i\sigma_0^y)  V^\dagger(u_{cd}) \ket{\upOmega}.
    \end{aligned}
\end{equation}
In the second line of the expression above, we have used the group homomorphism property of $V(u)$ under matrix multiplication, which allows us to combine the operators $V(u_t)$ and $V(u_{cd})$ as
\begin{equation}\label{eq: V_ut_compressed}
    \begin{aligned}
        V(u_t^{\prime}) &= V(u_t) \cdot  V(u_{cd}) \\
        &= V(u_t) \cdot \left( V^{\dagger}(\tilde{p}_0) V^{\dagger}(\tilde{r}_{N-1,0}) \cdots V^{\dagger}(\tilde{r}_{1,0}) \right) \cdot
        \left( V^{\dagger}(\tilde{p}_1) V^{\dagger}(\tilde{r}_{N-1,1}) \cdots V^{\dagger}(\tilde{r}_{2,1}) \right) \\
        &=V(u_t) \cdot \left( V(\tilde{p}_0^{\dagger}) V(\tilde{r}_{N-1,0}^{\dagger}) \cdots V(\tilde{r}_{1,0}^{\dagger}) \right) \cdot
        \left( V(\tilde{p}_1^{\dagger}) V( \tilde{r}_{N-1,1}^{\dagger}) \cdots V(\tilde{r}_{2,1}^{\dagger}) \right) \\
        &= V( u_t \cdot \tilde{p}_0^{\dagger} \cdot \tilde{r}_{N-1,0}^{\dagger} \cdots \tilde{r}_{1,0}^{\dagger} \cdot \tilde{p}_1^{\dagger} \cdot \tilde{r}_{N-1,1}^{\dagger} \cdots \tilde{r}_{2,1}^{\dagger} )
    \end{aligned}
\end{equation}
exploiting the property $V(u)^{\dagger} = V(u^{\dagger})$. Since the wave packets' coefficients $\tilde{\phi}_n^c$ and $\tilde{\phi}_n^{d\ast}$ are in the first and second column of $u_{cd}$, we should decompose the first two columns of $u_{cd}$ using Givens rotation, as described in Sec.~\ref{app:Givens_rotation}. In Eq.~\eqref{eq: V_ut_compressed}, $\tilde{p}_0, \tilde{p}_1$ and $\tilde{r}_{i,0}, \tilde{r}_{i,1}$ refer to the Givens rotation matrices required to diagonalize the first two columns of $u_{cd}$. Parallelizing the Givens rotations on distinct qubit-pairs, the time-evolution operator in the noninteracting case can be decomposed into a quantum circuit of constant depth  $\mathcal{O}(4N)$, using $\mathcal{O}(N^2)$ CNOT gates.

\begin{figure}
    \centering
    \includegraphics[width = 0.8\textwidth]{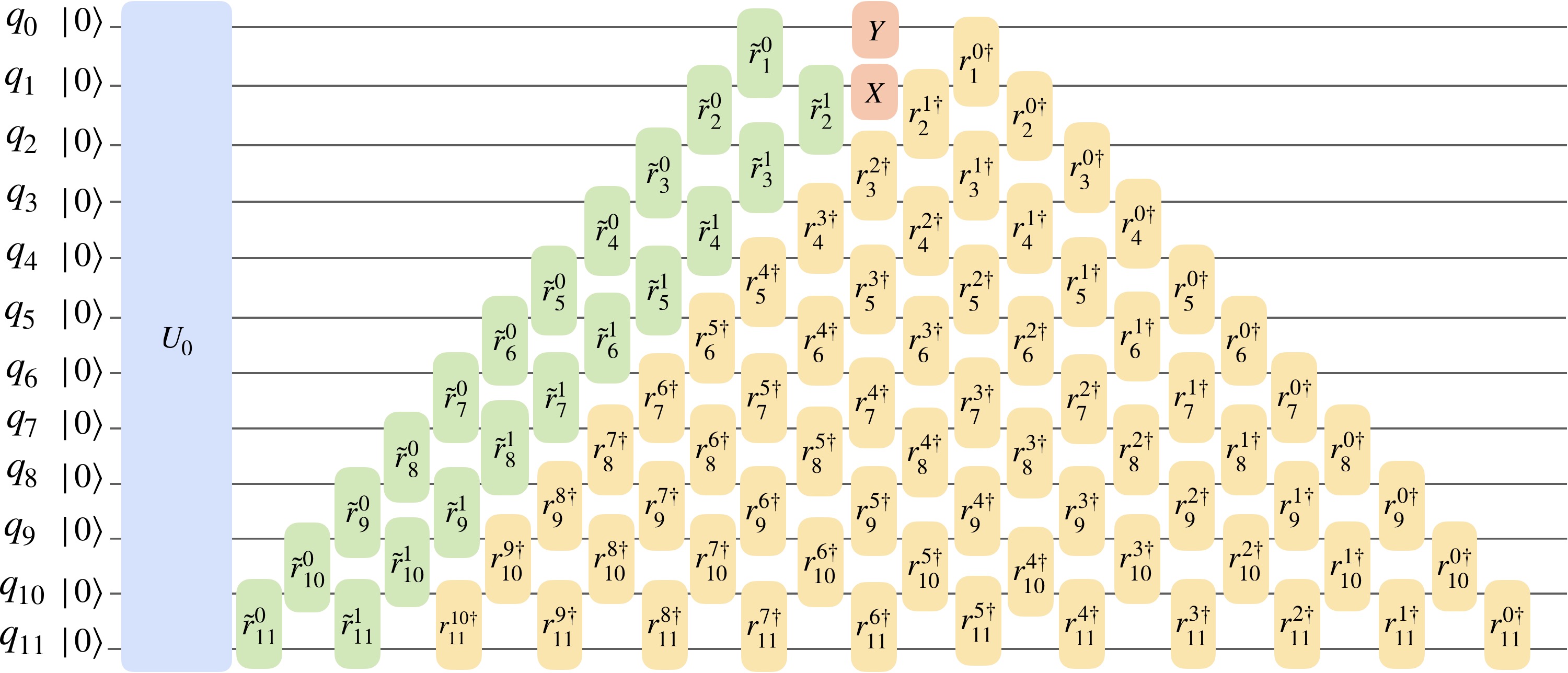}
    \caption{Circuit for the hardware run of free fermions. The blue box represents the circuit for vacuum preparation, where $U_0$ consists of $X$ gates on every second qubit, essentially preparing the state $\ket{01}^{\otimes N/2}$, followed by parametric iSwap gates. The additional $X$ gates are added because for $m=1$, the N\'eel state is expected to have a large overlap with the target ground state. As such, it serves as a suitable initial state for the VQE to find the ground state efficiently. By optimizing the parameters in $U_0$ through classical simulation, we can get the ground state with fidelity $99.7\%$ using 4 CNOT layers. The green and yellow boxes are circuits for the wave packet preparation and time evolution, each box represents a two-qubit gate $V(r)$ that consists of two CNOT gates as shown in Fig.~\ref{fig:givens}.}
    \label{fig:hardware_run_circuit}
\end{figure}

Finally, the state $\ket{\psi(t)}$ can be prepared using the quantum circuit in Fig.~\ref{fig:hardware_run_circuit}. The blue box represents the circuit $U_0$ for preparing the vacuum state; each green box corresponds to the Givens rotation gate $V(\tilde{r})$ for decomposing $V^{\dagger}(u_{cd})$, e.g.,
\begin{equation}
    \begin{aligned}
        V^{\dagger}(u_{cd}) &= V(\tilde{s}_1) \cdot V(\tilde{s}_0) \\
        &= \left( V(\tilde{r}_{2,1}) \cdot V(\tilde{r}_{3,1}) \cdots V(\tilde{r}_{N-1,1}) \cdot V(\tilde{p_1}) \right)
        \cdot \left( V(\tilde{r}_{1,0}) \cdot V(\tilde{r}_{2,0}) \cdots V(\tilde{r}_{N-1,0}) \cdot V(\tilde{p_0}) \right),
    \end{aligned}
\end{equation}
using the fact we can find the unitary matrices $\tilde{s}_0, \tilde{s}_1$ that consists of Givens rotation matrix as outlined in Eq.~\eqref{eq: s} and Eq.~\eqref{eq: s_rp}. Notice that the circuits for $V(\tilde{p}_0)$ and $V(\tilde{p}_1)$ correspond to single-qubit rotation gates (see Eq.~\eqref{eq: Vp_Rz}), and are not shown explicitly in Fig.~\ref{fig:hardware_run_circuit} for simplicity. The red boxes represent the Pauli operators $\sigma_0^y$ and $\sigma_1^x$ in Eq.~\eqref{eq: psit_free}, which will excite one fermion and one antifermion on the free fermion vacuum. The yellow boxes are the decomposition of the unitary operator $V(u_t^{\prime})$ implemented using its decomposition in Givens rotation matrices. 

In this work, the observable measured in the hardware run is the expectation value of the Pauli $Z$-operator, which corresponds to the particle density. For the measurement of qubits $n, (\ n = 0, 1, \dots 11$), operators in the circuit located outside the corresponding light cone do not influence the measured observables $\sigma^z_n$. For example, we can only keep the operators $V(r_{1,0}^{\dagger}) V(r_{2,1}^{\dagger}) \cdots V(r_{l+1,l}^{\dagger}) \cdots V(r_{11,10}^{\dagger})$ in the circuit for $V(u_t^{\prime})$ (yellow boxes in Fig.~\ref{fig:hardware_run_circuit}) when measuring $\sigma^z_0$, which reduce roughly half of the circuit depth for the wave packet preparation and time evolution. The circuit depth is 30 for measuring $\sigma_0^z$ (do not include the circuit $U_0$ for vacuum preparation and we only count the depth of the CNOT gates.), with an increasing circuit depth of 2 for one more qubit index in observable because of the expansion of the light cone. This results in a similar circuit as in Fig.~\ref{fig:hardware_run_circuit} but with an inverted triangle shape, which will result in a shallower circuit depth for the observables with a qubit index from 6 to 11. Thus, using the circuit in Fig.~\ref{fig:hardware_run_circuit} for measuring $\sigma^z_0, \dots \sigma^z_5$, and using the inverted triangle circuit for measuring $\sigma^z_6, \dots \sigma^z_{11}$, the circuit depth for wave packet preparation and time evolution has a range from 30 to 40, and 34 to 44 if including the circuit $U_0$ for vacuum preparation.

\section{Proof of orthogonality between fermion and antifermion wave packet coefficients}\label{app: orthogonality}
The normalized coefficients of fermion and antifermion wave packets in position space are defined by Eq.~\eqref{eq:coeff_real_sapce}, and the orthogonality of $\phi^c$ and $\phi^{d\ast}$ can be proven as follows:

\begin{equation}\label{eq: phi_c_d_ortho}
    \begin{aligned}
        &\sum_n \tilde{\phi}_n^c \times \tilde{\phi}_n^d = \frac{1}{N} \sum_{npq} \phi_p^c \phi_q^d \times \sqrt{\frac{m+w_p}{w_p}}\sqrt{ \frac{m+w_q}{w_q}} \times e^{in(p+q)} \times \left(\Uppi_{n0} \times v_q + \Uppi_{n1}\times v_p \right) \\
        &= \frac{1}{N} \sum_{pq} \phi_p^c \phi_q^d \times \sqrt{\frac{m+w_p}{w_p}}\sqrt{ \frac{m+w_q}{w_q}} \times \left( \frac{ \delta(p+q) + \delta(p+q-\pi) }{2} \times v_q +  \frac{ \delta(p+q) - \delta(p+q-\pi) }{2} \times v_p \right) \\
        &= \frac{1}{N} \sum_{p} \phi_p^c \phi_{-p}^d \times \sqrt{\frac{m+w_p}{w_p}}\sqrt{\frac{m+w_{-p}}{w_{-p}}} \times \left(v_{-p} + v_{p}\right) \\
        &+ \frac{1}{N} \sum_{p} \phi_p^c \phi_{\pi-p}^d \times\sqrt{\frac{m+w_p}{w_p}}\sqrt{ \frac{m+w_{\pi-p}}{w_{\pi-p}}} \times \left(v_{\pi-p} + v_{-p}\right)\\
        &= 0.
    \end{aligned}
\end{equation}
In the first line of the equation above, the coefficients are $\tilde{\phi}_n^{c(d)}$ defined in position space are represented by $\phi_k^{c(d)}$ defined in momentum space as in Eq.~\eqref{eq:coeff_real_sapce}. Besides, we use the property of projectors defined in Eq.~\eqref{eq:projector}, $\Uppi_{n0} \times \Uppi_{n1} = 0,\ \Uppi_{nl} \times \Uppi_{nl} = \Uppi_{nl}, \ l\in\{0, 1\}$. In the second line, the Fourier transformation on even and odd sites is performed:
\begin{equation}
    \begin{aligned}
        \sum_{n\in \text{even}} e^{in(p+q)} &= \frac{ \delta(p+q) + \delta(p+q-\pi) }{2}, \\
        \sum_{n\in \text{odd}} e^{in(p+q)} &= \frac{ \delta(p+q) - \delta(p+q-\pi) }{2}.
    \end{aligned}
\end{equation}
Recalling the definition of $v_p$ in Eq.~\eqref{eq:wk_vk}, we can get $v_{-p} = -v_p, v_{\pi - p} = - v_{-p}$, showing that Eq.~\eqref{eq: phi_c_d_ortho} results in zero. A similar procedure can be applied for the coefficients for two fermions or two antifermions, revealing that orthogonality does not exist in these cases.

\section{Qiskit primitives, error suppression and mitigation\label{app: qem}}
\begin{figure}
    \centering
    \includegraphics[width = 0.8\textwidth]{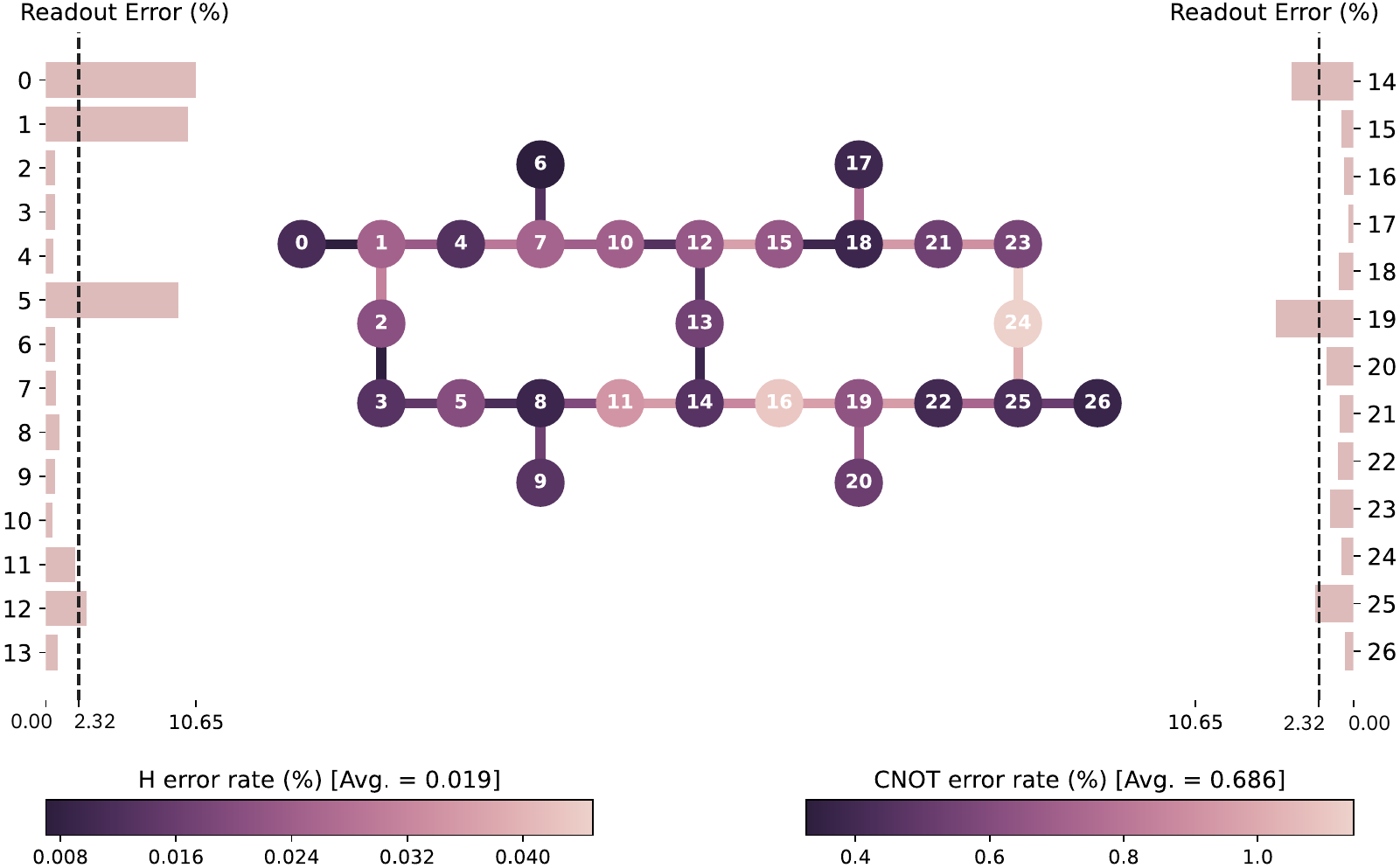}
    \caption{Error map for \texttt{ibm\_peekskill}, a Falcon R8 processor. Physical qubits used in this study are those from the unit cell containing qubit 1 due to having higher quality qubits and gates. The median $T_1$ and $T_2$ coherence times for these qubits were $335.81~\mu$s and $299.93~\mu$s, respectively. Mapping of logical to physical qubits took into account qubit and gate fidelity (see main text).}
    \label{fig:peekskill-error-map}
\end{figure}

All circuits were executed on \texttt{ibm\_peekskill}, a quantum device housing a 27-qubit IBM Falcon R8 processor. This backend has a heavy hexagonal topology comprising two unit cells whose common qubits are labeled 12, 13, and 14; see Fig~\ref{fig:peekskill-error-map}. In the studies presented, the unit cell containing qubit 1 was used due to having the best performing qubits and gates. To attain our results, we ensured that the most active qubits in our circuits were mapped to the qubits with the longest coherence times and highest fidelity two-qubit gates. This was achieved by using circuit-dependent permutations for logical-to-physical qubit mappings.

\begin{figure}[htp!]
    \centering     
    \includegraphics[width=\textwidth]{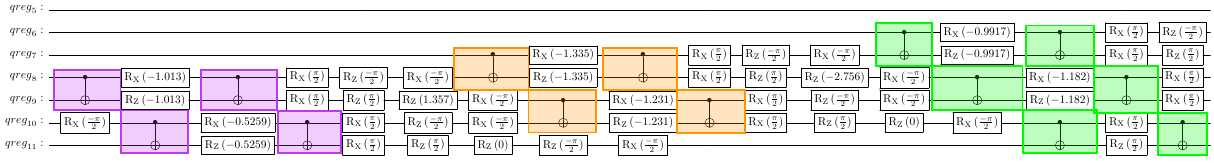}   \\ \vspace{1em}  
    \includegraphics[width=\textwidth]{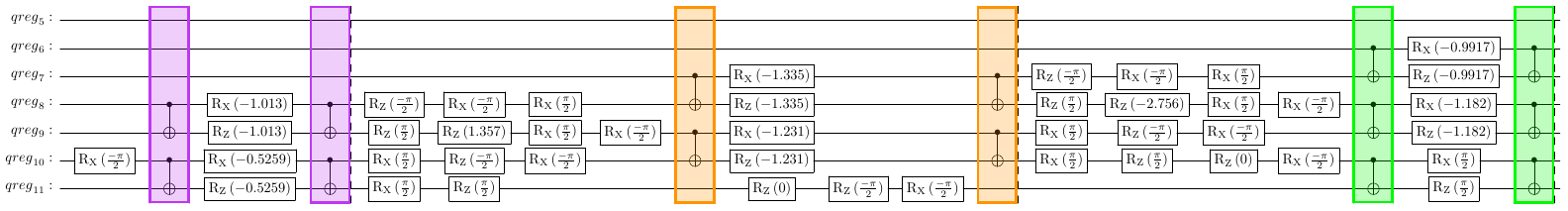}
    \caption{Subset of layers of one- and two-qubit gates from the $t=0$ circuit with idle wires removed, before transpilation and circuit optimization. Upper panel: the nominal staggered layout of entangling gates. Notice how the lack of structure due to offset CNOT gates. Lower panel: the restructured layout aligning entangling gates admitting fewer CNOT layers and therefore more efficient learning.}
    \label{fig:ex-circs}
\end{figure}

Each of the five circuits, corresponding to time evolution steps $t = \{0, 6, 12, 18, 24\}$, executed has a total depth of 174, 54 of which are entangling (two-qubit gate) layers. We note that unlike the highly structured circuits in Ref.~\cite{Kim2023}, our circuits do not comprise complete layers but rather partial layers of two-qubit gates.
That is, two-qubit entangling gates do not act in parallel across the width of the circuit but instead are staggered; see Fig.~\ref{fig:ex-circs}, upper panel. The implications, for example, include increased circuit execution time due to redundant learning when performing error mitigation (discussed below) and use of dynamical decoupling as most qubits are idle in these partial layers. To address this, we collected blocks of entangling gate layers to impose structure into our circuits such that these gates aligned to act in parallel, while also maintaining single-qubit gates ordering; see Fig.~\ref{fig:ex-circs}, lower panel. The added structure reduces the number of unique layers to learn by more than half, resulting in a similar decrease in overall circuit execution time.

\begin{figure*}
    \includegraphics[width=\linewidth]{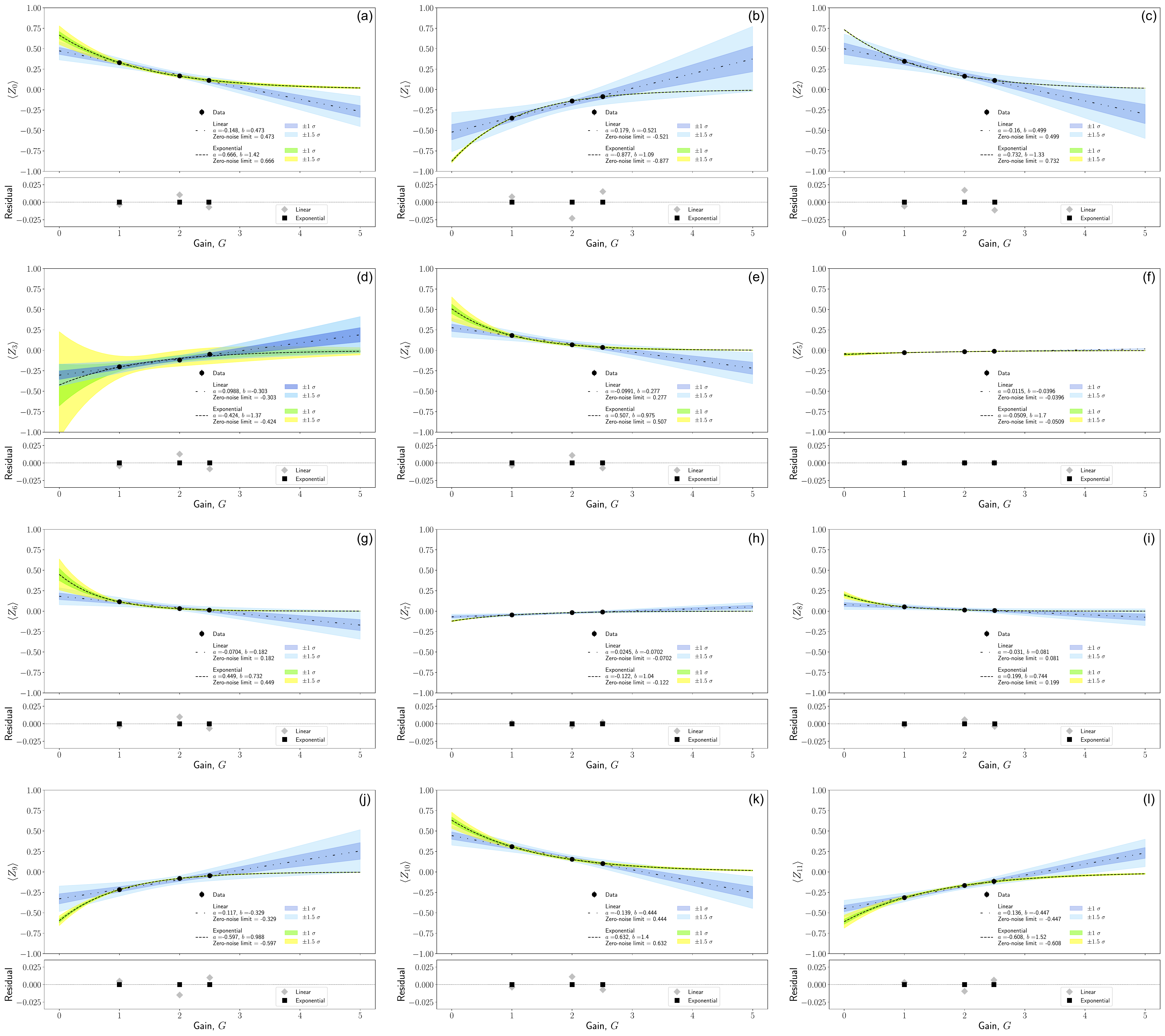}
    \caption{Example of zero-noise extrapolation (ZNE) of each observable $\langle Z_i \rangle$ at time $t=12$ using linear (dash-dot) and exponential (dash-dash) models, along with their respective $1~\sigma$ and $1.5~\sigma$ uncertainty bands. The choice of fit was determined by the model providing the smallest residuals overall. Data points are shown in solid dots. Gains $G = \{1, 2, 2.5\}$ were chosen such that circuit depths remain within device coherence times and signal decayed with increasing gain. Extrapolated estimates of each observable at each time step $t$ were transformed into the corresponding fermion charge density via $0.5 \times (1.0 - \langle Z \rangle)$ and subtracting the vacuum charge density.}
\label{fig:zne_plots}
\end{figure*}

In order to mitigate the effects of hardware noise, we systematically increase the noise in our circuits using probabilistic error amplification (PEA)~\cite{Kim2023} and determine an estimate of the zero-noise limit with zero-noise extrapolation~\cite{PhysRevLett.119.180509, Li2017}. In PEA, we apply Pauli twirling to unique layers of entangling gates, $l$, to shape the noise, modelled as sparse Pauli-Lindblad noise channels~\cite{vandenBerg2023},
\begin{equation}
    \Uplambda_l =
    \prod_{k \in \mathcal{K}} \left( w_k \cdot + (1-w_k)P_k  P_k^{\dagger} \right) \rho,
\end{equation}
where $\mathcal{K}$ is a set of local Pauli ``model terms'' capturing noise interactions in the quantum processor, $w_k = (1 + e^{-2 \lambda_k})/2$, and $\lambda_k$ is the corresponding model coefficient.
Learning circuits are constructed from each $l$ by repeating the layer $\{0, 4, 16\}$ times, each depth sampled 16 times.
In addition, we employ readout-error mitigation by applying a Pauli $X$ or $I$ operator~\cite{PhysRevA.105.032620}, sampled pseudo-randomly, to precede measurements.
The $X_p - X_m$ dynamical decoupling is applied to idle qubits that are inactive within entangling layers.
We generated 300 circuit instances at gain factors $G = \{1.0, 2.0, 2.5\}$, corresponding to scaled noise channels, $\Uplambda_l^{G-1}$, and sampled each of the resulting 900 circuits 1024 times to determine an estimate of the expectation value for each observable for each $G$.
Extrapolation to the zero-noise limit was performed using the averaged expectation value over all instances at each $G$ using linear and exponential fits having accepted fit parameters with the smallest least squares residual. Figure~\ref{fig:zne_plots} shows our fits of all observables $\langle Z_i \rangle$ for time $t=12$.

This work leveraged the latest offerings from the Qiskit Primitives API~\cite{QiskitPrimitives}. In particular, we utilized the \texttt{Estimator} which returns an estimate for an observable's expectation value. This relieves one of dealing directly with individual measurements through its mechanisms for calculating expectation values of observables. 
In this work, we execute the hardware run for five-time slices, and there are 12 distinct circuits for observables $\sigma_n^z$ on different sites, which only include the gates inside the light cone as detailed in Sec.~\ref{app: decompose_Ut_Givens}. Each circuit -- after applying error mitigation and suppression, see above -- is executed sequentially on the device, the corresponding observable measured, and a \texttt{Job} object is returned containing the execution results. In all, we have $60 \times 900=54000 $ circuits, and $1024$ measurement shots for each circuit.

\twocolumngrid
\FloatBarrier
\bibliographystyle{quantum}
\bibliography{references}

\end{document}